\documentclass[pra,amsmath,aps,amssymb,nofootinbib,superscriptaddress,floatfix,twocolumn]{revtex4-2}
\usepackage[mathscr]{euscript} 
\usepackage{graphicx}
\usepackage{upgreek}
\usepackage{dcolumn}
\usepackage[colorlinks=true,linkcolor=blue,urlcolor=blue,citecolor=blue]{hyperref}
\usepackage{float}
\usepackage{caption}
\usepackage{subcaption}
\usepackage{orcidlink}
\usepackage{mathtools}
\usepackage{epsfig}
\usepackage{epstopdf} 
\usepackage{siunitx}
\usepackage{booktabs}
\usepackage{makecell}
\usepackage{multirow}
\begin{document}

\preprint{APS/123-QED}

\title{Calibration Method of Spacecraft–Inertial Sensor Center-of-Mass Offset for the Taiji Gravitational Wave Detection Mission under Science Mode}

\author{Haoyue Zhang}
\affiliation{State Key Laboratory of Micro-Spacecraft Rapid Design and Intelligent Cluster, Harbin Institute of Technology, Harbin 150001, China}%
\affiliation{Center for Gravitational Wave Experiment, Institute of Mechanics, Chinese Academy of Sciences, Beijing 100190, China}

\author{Dong Ye}%
\email{yed@hit.edu.cn}
\affiliation{State Key Laboratory of Micro-Spacecraft Rapid Design and Intelligent Cluster, Harbin Institute of Technology, Harbin 150001, China}%

\author{Peng Xu}%
\email{xupeng@imech.ac.cn}
\affiliation{Center for Gravitational Wave Experiment, Institute of Mechanics, Chinese Academy of Sciences, Beijing 100190, China}%
\affiliation{Hangzhou Institute for Advanced Study, University of Chinese Academy of Sciences, Hangzhou 310024, China}%
\affiliation{Lanzhou Center of Theoretical Physics, Lanzhou University, Lanzhou 730000, China}

\author{Yunhai Geng}%
\affiliation{State Key Laboratory of Micro-Spacecraft Rapid Design and Intelligent Cluster, Harbin Institute of Technology, Harbin 150001, China}%

\author{Li-E Qiang}%
\affiliation{National Space Science Center, Chinese Academy of Sciences, Beijing 100190, China}%

\author{Ziren Luo}%
\affiliation{Center for Gravitational Wave Experiment, Institute of Mechanics, Chinese Academy of Sciences, Beijing 100190, China}%

\date{\today}

\begin{abstract}

Accurately calibrating the center-of-mass (CoM) offset between the spacecraft (SC) and the inertial sensor test mass (TM) is crucial for space-based gravitational wave (GW) antennas, such as LISA and Taiji. Current calibration methods require additional spacecraft maneuvers that disrupt science data continuity and inter-satellite links, compromising the coherence of GW signals. Here, we present a maneuver-free calibration scheme that directly estimates the CoM offset vector using only standard science-mode measurements from inertial sensors, interferometers, and differential wavefront sensors. By embedding the CoM offset induced acceleration as an extended state in a model-based adaptive Kalman filter, we achieve estimation accuracy of 0.01–1.5 mm across all axes with a maximum error below 1\%. This approach enables continuous, high-precision calibration during nominal science operations, ensuring continuous and coherent gravitational wave data collection while maintaining the required precision, and also facilitating advanced DFACS functions such as performance evaluations and fault diagnosis. For LISA-like missions, where data continuity is paramount for detecting faint and long-standing GW signals, this method will enhance scientific output and reliability.

\end{abstract}

\maketitle


\section{\label{sec1}Introduction}

The Laser Interferometer Space Antenna (LISA) mission \cite{LISA2024} was officially approved by the European Space Agency in January 2024. As the most developed space mission, LISA had motivated various space gravitational wave (GW) detection projects since the 2000s, including DECIGO \cite{DECIGOsato2017status,DECIGOando2010decigo,DECIGOkawamura2011japanese}, Taiji \cite{TAIJI_OVERVIEW1,hu2017taiji}, TainQin \cite{TQ_OVERVIEW,TQ_liu2020science,TQ_mei2021tianqin}, ASTROD \cite{ni2002astrod,ni2008astrod} and etc. Compared with the ground-based large interferometers, that the LIGO-Virgo-KAGRA collaboration \cite{LIGO2016,caron1997virgo,kagra_sekiguchi2012current,lisa_virgo_kagra_abbott2020prospects}, space-borne antennas will be free from Newtonian gradient noises situated near the 10 Hz band and limitations on the baselines. The LISA and the mentioned LISA-like missions will have sensitive frequency bands from 0.1 mHz to 1 Hz and will encompass scientific objectives related to much heavier sources such as coalescing super-massive black holes and extreme-mass-ratio inspirals, and also stochastic GW background from the early Universe. 

``The Taiji program in space'' is a space-borne GW detection mission of China, released by the Chinese Academy of Sciences in 2016. Following its roadmap \cite{luo2020taiji}, the first technology demonstration satellite, that the Taiji-1, was launched in August 2019 and has successfully validated the related key technologies including the high precision inertial sensor (IS, also called GRS, Gravitational Reference Sensor)~\cite{GRS_2022}, laser interferometer in space, drag-free controls and ultra-clean and stable satellite platform \cite{TAIJI-1_OVERVIEW}. The Taiji mission, as one of the representative LISA-like missions, will launch three spacecraft (SC) to heliocentric orbits, forming a nearly equilateral triangular constellation with armlength $\sim 3\times 10^6$ km. It is expected that the science operations of LISA and Taiji may overlap in the 2030s and studies of the joint data analysis has already aroused more interest \cite{lisataiji_wang2021alternative,lisataiji_cornish_shuman2022massive,LISATAIJI_ruan2020lisa,LISATAIJI_orlando2021measuring}.


High precision IS are of key payloads for the LISA and Taiji missions. 
The suspended test mass (TM) of each IS serves as the free-falling reference and also as the end mirror for the inter-satellite interferometers. 
The relative motions between the TMs induced by incident GWs could be measured precisely by means of the so called time-delay interferometry \cite{TDI_tinto2021time} and therefore give rise to the detection of GW signals.
The performance of the IS, or the residual acceleration noises of the TMs relative to the local inertial frame, will directly determine the sensitivity level of the space GW antennas.
For the Taiji mission, according to its designed sensitivity and noise budgets, the residual acceleration noise for each TM along the sensitive axis (along with the inter-satellite laser link) of the IS should be $\leq 3\times 10^{-15}\ \mathrm{m/s^{2}/Hz^{\frac{1}{2}}}$ ranging from 0.1 mHz
to 1 Hz.~\cite{luo2020brief}. 
Such stringent requirement will be upheld and guaranteed by the Drag-Free and Attitude Control System (DFACS)~\cite{bender2000lisa}, which monitors in real-time and adjusts actively the relative displacements and attitudes between the SC and the TMs. 
The DFACS maintains the two TMs on-board each SC in nearly perfect free fall along their sensitive axes, while forcing them to follow the SC for the remaining degrees of freedom (including attitudes)~\cite{gath2004drag}. 
Given the careful control of each SC’s attitude and the breathing angle of the movable optical sub-assemblies to ensure stable inter-satellite measurement links,  the attitude of the TMs will be maintained to align with the laser links, effectively acting as the free-falling end mirrors of the inter-satellite interferometer.

Extensive studies on the DFACS for LISA-like missions have established a comprehensive model of the relative dynamics of the TMs and SC~\cite{gath2007drag}. Within this model, the relative center-of-mass (CoM) offset between the SC and TM (called the SC–TM CoM offset) constitutes the most significant nonlinear effect and gives rise to an important disturbance to the DFACS. 
The CoM offset vector and its variations will couple with the angular acceleration and angular velocity of the SC respectively, thereby introducing inertial acceleration noises in the local frame attached to the SC. 
Such CoM coupled noise will also introduce interference into the performance evaluation of the on-board ISs.  
For example, in the data analysis work of the LISA Pathfinder, the CoM offset was precisely calibrated, and its influence, that the inertial force noises, was accurately removed in obtaining the expected noise curve \cite{LPF2018calibrating}.

For satellite gravity missions, such as CHAMP \cite{reigber2002champ}, GRACE \cite{GRACE_weil1997gravity}, and GFO \cite{gfo_kornfeld2019grace}, the onboard IS operated in accelerometer mode, and their readouts provided the necessary measurements of non-gravitational perturbations acting on the corresponding SCs.  
The CoM coupled effects will produce inertial acceleration noises directly to the measured data, degrading the scientific outputs of satellite missions with accelerometers for the measuring non-gravitational orbit perturbations.
Therefore, the in situ and accurate calibration of the SC–TM CoM offset is of importance~\cite{COM_CALI_TH} for those missions also including GW antennas in space. 
Take GRACE as a representative example, one performed the CoM calibration by commanding large spacecraft attitude maneuvers with magnetorquers. 
The attitude jitters and the CoM coupled acceleration signals were measured with star-trackers and the accelerometer, and then the offset vector was estimated using the least-squares method~\cite{wang2003study,COM_CALI_WANG_JOURNAL}. 
Taiji-1 ~\cite{taiji2021china,COM_CALI_ZHANG,COM_CALI_WEI} and TianQin-1 (the technology demonstration satellite of the TianQin project)~\cite{tq1_zhou2022non,TQ1_RESULT} adopted the similar maneuver strategy and employed Kalman filtering algorithms to estimate the CoM offset. 
The feasibility of applying this method to the Taiji mission was studied and verified with numerical simulations \cite{liu2025design}.
Furthermore, the TianQin team also analyzed the maneuver design criteria for the possible CoM calibration, which could maintain the inter-satellite link at the same time~\cite{wang2025research}. 
In short, during their commissioning and calibration phases, the maneuver-based SC–TM CoM offset calibration methods could meet the requirements of those LISA-like missions~\cite{lisa2011lisa,liu2025design}.

During the years-long science lifetime of Taiji and LISA, the SC-TM CoM offset vector could change slowly with time, which may be caused by the continuous consumption of the propellant, the structural deformation due to material aging, or the possible drift of the balance point of the TMs. 
Therefore, the scheduled CoM calibration in the science operation phase will be necessary in ensuring the quality of scientific data and the accurate assessment of the performance of the ISs themselves \cite{lupi2019precise}. 
However, the maneuver-based calibration will cost extra price when the space-borne GW antenna is performing science measurements.
The rather large attitude swings (such as the 100 \(\mathrm{\mu N\cdot m}\) maneuver proposed in~\cite{wang2025research}) would consume a substantial portion of the DFACS control margin and severely compromise the stability of the inter-spacecraft laser links. 
This contradicts the very purpose of using the \textit{H$_\infty$} control approach in DFACS to ensure system robustness. 
More critically, the maneuvers and the resulting coupled acceleration noises would disrupt the continuity of the science data and break the coherence of the GW signals, thereby would increase the difficulty of source parameter estimations and may even affect the final scientific output of the missions.

To meet the CoM offset calibration requirements of the Taiji mission, this paper proposes a maneuver-free scheme applicable during the science run without any disruption to the measurement system. 
As discussed earlier, the CoM offset induces disturbances that can substantially influence the performance of the DFACS system, which, correspondingly, will be manifested in the readout signals of the DFACS measurement subsystems—namely, the local TM laser interferometers (IFO), differential wavefront sensors (DWS)~\cite{OB_brzozowski2022lisa}, and ISs. These readouts and their correlations make it possible to identify the SC–TM CoM offset vector given the SC–TM dynamical model.
We developed a high-fidelity DFACS numerical simulations
and generate the synthetic readout data set. By embedding the CoM offset as an extended state, alongside the bias terms and the dominant colored noises, we reformulate this problem as an extended estimation problem that could be solved via an improved Sage--Husa adaptive Kalman filter. The proposed method enables high-precision CoM offset identification directly from undisrupted science-mode data, without requiring auxiliary maneuvers.

This paper expands as follows. Section~\ref{sec2} introduces the purpose of the DFACS system, describes the physical configuration and operating principle of the IS, and presents the equations governing the SC–TM relative dynamics and spacecraft attitude motion. Section~\ref{sec3} designs a DFACS for Taiji and provides a high-precision simulation of the relative attitude/displacement between the SC-TM and the SC attitude signal in the inertial frame under science mode. Section \ref{sec4} presents the CoM calibration methodology, detailing a novel maneuver-free approach that relies solely on DFACS-provided SC–TM relative displacement data. The method is validated using 25 days of high-fidelity DFACS simulation data, and the resulting CoM offset estimates are reported. Finally, Section~\ref{sec5} concludes the paper and discusses implications for mission operations.

\section{\label{sec2}Spacecraft-Test Mass Dynamics Model}
\begin{figure}[htbp]
\centering
\includegraphics[width=0.95\columnwidth]{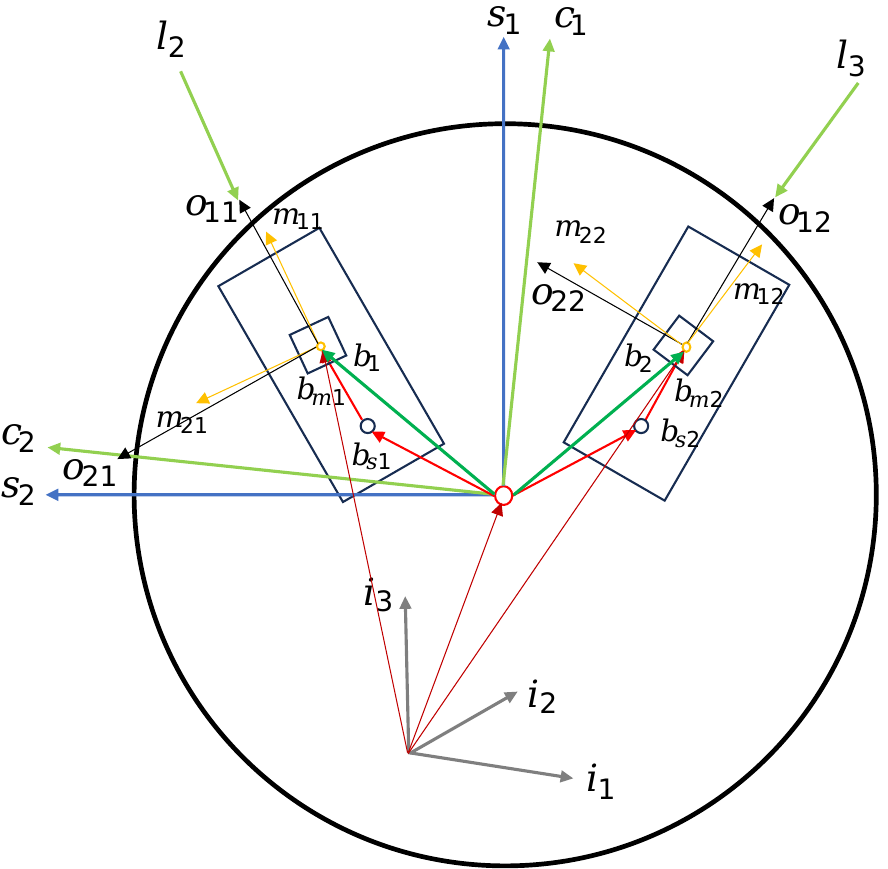}
\caption{Reference coordinate system definitions and labeling of key vectors for SC1.}
\label{fig:epsart}
\end{figure}

Fig.~\ref{fig:epsart} illustrates the internal layout configuration of the Taiji spacecraft. For concreteness, SC1 is used as a representative example. Inside SC1, two TMs undergo free fall along their respective sensing axes, $\mathbf{o}_{11}$ and $\mathbf{o}_{12}$, which are separated by an angle of approximately $60^\circ$. Each TM is enclosed within an electrode cage, rigidly mounted to its associated optical bench and to the SC platform. The IS employs sensing electrodes in the cage to measure the TM-SC relative displacement and rotation, and actuation electrodes to apply electrostatic forces on the TM. 

Owing to their shared measurement principle, LISA, Taiji, and TianQin feature highly similar internal mechanical layouts~\cite{DFACSbohan2024review}. we adopt the coordinate systems established in~\cite{vidano2020lisa}, which provide consistent frames for expressing forces, torques, and kinematic relationships. The key reference frames illustrated in Fig.~\ref{fig:epsart} are defined as follows:
\begin{itemize}
    \item Heliocentric Inertial Reference Frame (IRF): 
    $\displaystyle \text{IRF} = \{ O^I,\, i_1,\, i_2,\, i_3 \}$,
    \item Constellation Reference Frame (CRF): 
    $\displaystyle \text{CRF} = \{ O^C,\, c_1,\, c_2,\, c_3 \}$, $\mathbf{c}_1$ is aligned with the bisector of the incoming laser beam directions $\mathbf{l}_2$ (from SC2) and $\mathbf{l}_3$ (from SC3); $\mathbf{c}_2$ lies in the plane spanned by $\mathbf{l}_2$ and $\mathbf{l}_3$, orthogonal to $\mathbf{c}_1$, $\mathbf{c}_3 = \mathbf{c}_1 \times \mathbf{c}_2$.
    \item Spacecraft Reference Frame (SRF): 
    $\displaystyle \text{SRF} = \{ O^S,\, s_1,\, s_2,\, s_3 \}$.
    \item Optical Reference Frame of the $i$-th optical bench (ORF$_i$): 
    $\displaystyle \text{ORF}_i = \{ O^{O_i},\, o_{1i},\, o_{2i},\, o_{3i} \}$, for $i = 1,2$,
    \item Test-Mass Reference Frame of the $i$-th TM (MRF$_i$):
    $\displaystyle \text{MRF}_i = \{ O^{m_i},\, m_{1i},\, m_{2i},\, m_{3i} \}$, for $i = 1,2$.
\end{itemize}
Specifically, the superscript of a variable or parameter denotes the reference frame in which it is defined, while the subscript indicates the relative relationship or the entities involved in that variable. For example:\\
$\mathbf{r}_{m_i O_i}^{O_i}$: The relative displacement between the i-th optical bench and the i-th TM in \(\text{ORF}_i\).\\
$\mathbf{b}_i$: The center-of-mass offset vector of the i-th test mass relative to the CoM of SC. \\
\(\mathbf{q}_{SI}, \mathbf{q}_{SC}\): Attitude quaternions of the SC relative to the IRF and CRF, respectively.\\
\(\mathbf{q}_{m_iS}\): Attitude quaternions of the i-th TM relative to the SC.\\
\(\mathbf{T}_A^B\): The Direction Cosine Matrix (DCM) that represents the coordinate transformation from frame A to frame B.


As the rigorous derivations of DFACS dynamics have been well established in previous studies (e.g., \cite{vidano2020lisa,DFACSbohan2024review}), this paper focuses on the key dynamic relationships essential for controller design and analysis.
Specifically, we present: 
(i) the translational and rotational dynamics of the $i$-th TM relative to the \(\text{ORF}_i\) — namely, $\mathbf{r}_{m_i O_i}^{O_i}$ and $\mathbf{q}_{m_i O_i}$ — under drag-free control; 
and (ii) the attitude dynamics of the SC relative to the CRF, described by $\mathbf{q}_{SC}$, under attitude control. 

Accurate modeling of the relative dynamics between the SC and its free-falling TMs is fundamental to the DFACS and CoM calibrations for Taiji.  To formulate these dynamics, we begin with the translational dynamics of the SC's CoM in the IRF:

\begin{equation}
\ddot{\mathbf{r}}_{SI}^I = \mathbf{a}_g(\mathbf{r}_{SI}^I) + m_S^{-1} \mathbf{T}_S^I \left( \mathbf{F}_T + \mathbf{d}_S - \sum_{i=1}^{2} \mathbf{T}_{O_i}^S \mathbf{F}_{E_i} \right).
\label{d2r_SC_I}
\end{equation}
Here, \( m_S \) is the mass of SC, \( \mathbf{F}_T \) is the thrust of the micro-propulsion system (MPS), \( \mathbf{d}_S \) represents environmental forces, \( \mathbf{F}_{E_i} \) is the electrostatic
forces used to control the TM, and \( \mathbf{a}_g(\mathbf{r}_{SI}^I) \) is the gravitational acceleration experienced by the SC, given by
\begin{equation}
\mathbf{a}_g(\mathbf{r}) = 
-\frac{\mu_s}{\|\mathbf{r} - \mathbf{r}_{\text{sun}}\|^3} (\mathbf{r} - \mathbf{r}_{\text{sun}})
-\frac{\mu_e}{\|\mathbf{r} - \mathbf{r}_e\|^3} (\mathbf{r} - \mathbf{r}_e),
\label{a_g}
\end{equation}
where $\mathbf{r}_{\text{sun}}$ and $\mathbf{r}_{e}$ are the position vectors of the Sun and Earth’s mass centers in IRF. \( \mu_s \) is the solar gravitational constant, \( \mu_e \) is the Earth's gravitational constant. Similarly, the acceleration of TM in IRF is given by
\begin{equation}
\ddot{\mathbf{r}}_{mI}^I = \mathbf{a}_g(\mathbf{r}_{mI}^I) + m_m^{-1} \, \mathbf{T}_O^I \left( \mathbf{F}_E + \mathbf{F}_{\mathrm{st}} + \mathbf{d}_m \right),
\label{eq:TM_acceleration_inertial}
\end{equation}
where \( \mathbf{a}_g(r_m^I) \) is the gravitational acceleration experienced by TM, \( \mathrm{m_m} \) is the mass of TM, \( \mathbf{F}_{\mathrm{st}} \) represents the TM stiffness force, primarily induced by the self-gravity gradient~\cite{stiffness_2009Current}. \( \mathbf{d}_m \) denotes other environmental noise on the TM. Finally, the relative acceleration of the i-th TM with respect to the SC can be expressed as
\begin{equation}
\ddot{\mathbf{r}}_{m_i O_i}^{O_i} = \mathbf{T}_I^{O_i} \left( 
    \ddot{\mathbf{r}}_{m_i I}^I 
    - \ddot{\mathbf{r}}_{SI}^I 
    - \mathbf{T}_S^I \, \boldsymbol{\Omega}(\boldsymbol{\omega}_{SI}) \, \mathbf{b}_i^S 
\right),
\label{eq:TM_SC_relative_acceleration}
\end{equation}
where 
\[
\boldsymbol{\Omega}(\boldsymbol{\omega}_{SI}) \coloneqq \boldsymbol{\omega}_{SI}^\times \boldsymbol{\omega}_{SI}^\times + \dot{\boldsymbol{\omega}}_{SI}^\times.
\]
$\boldsymbol{\omega}_{SI}^\times$ denotes the skew-symmetric matrix of $\boldsymbol{\omega}_{SI}$ (i.e., $\boldsymbol{\omega}_{SI}^\times \mathbf{v} = \boldsymbol{\omega}_{SI} \times \mathbf{v}$), and $\mathbf{b}_i^S$ is the i-th SC-TM CoM offset expressed in the SRF.

In Eq.~\eqref{eq:TM_SC_relative_acceleration}, $\boldsymbol{\omega}_{SI}$ represents the angular velocity of the SC relative to the IRF expressed in SRF. 
Unless otherwise specified, all angular velocities and angular accelerations (e.g., $\boldsymbol{\omega}_{SI}$, $\dot{\boldsymbol{\omega}}_{SI}$) are expressed in their body fixed reference frame; thus, explicit superscripts (e.g., ${}^S$) are omitted for brevity throughout this work.


The attitude dynamic of the SC relative to the IRF is given by
\begin{align}
\dot{\boldsymbol{\omega}}_{SI} &= \mathbf{J}_S^{-1} \left( \boldsymbol{\omega}_{SI} \times \mathbf{J}_S \boldsymbol{\omega}_{SI} \right) \notag \\
&\quad + \mathbf{J}_S^{-1} \Bigg( 
    \mathbf{M}_T + \mathbf{D}_S 
    - \sum_{i=1}^{2} \mathbf{T}_{O_i}^S \mathbf{M}_{E_i}\notag\\ 
    &\quad+ \sum_{i=1}^{2} \mathbf{b}_i^S \times \left( \mathbf{T}_{O_i}^S \mathbf{F}_{E_i} \right)
\Bigg),
\label{eq:SC_attitude_dynamics}
\end{align}
where $\mathbf{J}_S$ is the inertia moment of the SC, 
$\mathbf{M}_T$ is the control torque generated by the MPS, 
$\mathbf{D}_S$ denotes the environmental disturbance torque, 
and $\mathbf{M}_{E_i}$ represents the electrostatic torque applied on the $i$-th TM. 
Based on Eq.~\eqref{eq:SC_attitude_dynamics}, the angular acceleration of the SC relative to the CRF can be obtained as
\begin{align}
\boldsymbol\omega_{SC} &= \boldsymbol\omega_{SI} - \mathbf{T}_C^{S} \boldsymbol\omega_{CI}
\label{eq:SC_angular_velocity_CRF}, \\
\dot{\boldsymbol{\omega}}_{SC} &= 
\dot{\boldsymbol{\omega}}_{SI} 
+ \boldsymbol{\omega}_{SC}^\times \left( \mathbf{T}_C^S \boldsymbol{\omega}_{CI} \right)
- \mathbf{T}_C^S \dot{\boldsymbol{\omega}}_{CI},
\label{eq:SC_angular_acceleration_CRF}
\end{align}
where \( {\boldsymbol{\omega}}_{CI} \) is the angular velocity of the CRF relative to the IRF. This angular velocity is obtained by
\begin{equation}
\boldsymbol\omega_{CI}^\times = \mathbf{T}_I^C \dot{\mathbf{T}}_C^I.
\label{eq:CRF_angular_velocity}
\end{equation}
Furthermore, its angular acceleration is given by
\begin{equation}
\dot{\boldsymbol\omega}_{CI}^\times  = \mathbf{T}_I^C \left( \ddot{\mathbf{T}}_C^I - \dot{\mathbf{T}}_C^I \boldsymbol\omega_{CI}^\times \right).
\label{eq:CRF_angular_acceleration}
\end{equation}
Taking SC1 as the reference, let $\mathbf{r}_{12}^I$ and $\mathbf{r}_{13}^I$ denote the position vectors from SC1 to SC2 and SC1 to SC3, respectively, expressed in the IRF. 
Define the unit vectors
\begin{align*}
\hat{\mathbf{r}}_{12} &= \frac{\mathbf{r}_{12}^I}{\|\mathbf{r}_{12}^I\|}, &
\hat{\mathbf{r}}_{13} &= \frac{\mathbf{r}_{13}^I}{\|\mathbf{r}_{13}^I\|}, \\
\hat{\mathbf{x}}_C^I &= \frac{\hat{\mathbf{r}}_{12} + \hat{\mathbf{r}}_{13}}{\|\hat{\mathbf{r}}_{12} + \hat{\mathbf{r}}_{13}\|}, &
\hat{\mathbf{z}}_C^I &= \frac{\mathbf{r}_{12}^I \times \mathbf{r}_{13}^I}{\|\mathbf{r}_{12}^I \times \mathbf{r}_{13}^I\|}, \\
\hat{\mathbf{y}}_C^I &= \hat{\mathbf{z}}_C^I \times \hat{\mathbf{x}}_C^I,
\end{align*}
where $\hat{\mathbf{x}}_C^I$ aligns with the bisector of the formation (outward-pointing), $\hat{\mathbf{z}}_C^I$ is normal to the orbital plane, and $\hat{\mathbf{y}}_C^I$ completes the right-handed orthonormal basis. 
Then, $\mathbf{T}_C^I$ is given by
\begin{equation}
\mathbf{T}_C^I = 
\begin{bmatrix}
\hat{\mathbf{x}}_C^I, & \hat{\mathbf{y}}_C^I, & \hat{\mathbf{z}}_C^I
\end{bmatrix}.
\label{eq:TCI_transformation}
\end{equation}

The attitude dynamic of the TM relative to the IRF is given by
\begin{equation}
\dot{\boldsymbol{\omega}}_{m I} = 
\mathbf{J}_{m}^{-1} \left( \boldsymbol{\omega}_{m I} ^\times \mathbf{J}_{m} \boldsymbol{\omega}_{m I} \right)
+ \mathbf{J}_{m}^{-1} \left( \mathbf{M}_{E} + \mathbf{D}_{m} + \mathbf{M}_{\mathrm{st}} \right),
\label{eq:TM_attitude_dynamics}
\end{equation}
where \( \mathbf{J}_{m} \) is the inertia matrix of TM, \( \mathbf{M}_{E_i} \) is the electrostatic control torque (appears in Eq.~\eqref{eq:SC_attitude_dynamics}),  \( \mathbf{D}_{m_i} \) denotes environmental disturbance torques, and \( \mathbf{M}_{\mathrm{st}} \) is the stiffness torque on the TM~(analogous to the structural force $\mathbf{F}_{\mathrm{st},i}$ in Eq.~\eqref{eq:TM_acceleration_inertial}). Based on Eq.~\eqref{eq:TM_attitude_dynamics}, the angular velocity and angular acceleration of the i-th TM relative to the SC are obtained as
\begin{align}
\boldsymbol\omega_{m_i O_i} &= \boldsymbol\omega_{m_i I} - \mathbf{T}_S^{m_i} \boldsymbol\omega_{SI},
\label{eq:TM_angular_velocity_relative} \\
\dot{\boldsymbol\omega}_{m_i O_i} &= \dot{\boldsymbol\omega}_{m_i I} + \boldsymbol\omega_{m_i O_i}^\times \mathbf{T}_S^{m_i} \boldsymbol\omega_{SI} - \mathbf{T}_S^{m_i} \dot{\boldsymbol\omega}_{SI}.
\label{eq:TM_angular_acceleration_relative}
\end{align}

\section{\label{sec3}Drag-free and attitude control Design and Simulation Results}
\subsection{\label{subsec:Controller design}Controller design}

To ensure robust performance when confronted with environmental disturbances and model uncertainties during science operations, an $H_\infty$ control framework is employed for DFACS synthesis. This approach has been extensively considered or applied to different space-borne GW mission concepts~\cite{gath2004drag,gath2007drag}\cite{yidi2023robust}\cite{xu2024finite}. Its performance has been conclusively demonstrated by LISA Pathfinder~\cite{LPF2018calibrating} which achieved drag-free and attitude control performance at the nm/$\sqrt{\mathrm{Hz}}$ and nrad levels, respectively.

\begin{figure}[htbp]
\includegraphics[scale=0.6]{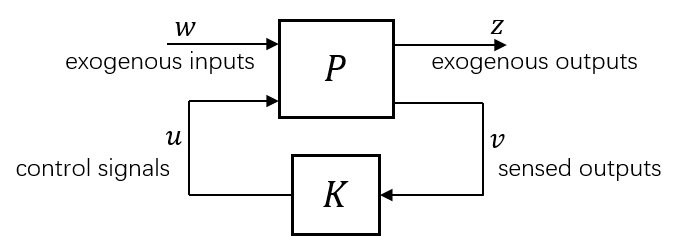}
\centering
\caption{\label{fig:H_inf} Generalized \(H_{\infty}\) control loop.}
\end{figure}

The $H_\infty$ control method is a robust synthesis approach rooted in norm-optimal control theory, specifically formulated to ensure closed-loop stability and prescribed performance bounds in the presence of model uncertainties and exogenous disturbances. As illustrated in Fig.~\ref{fig:H_inf}, the generalized plant $P$ represents the controlled dynamics (e.g., SC–TM coupled system), while $K$ denotes the to-be-designed controller. The objective is to synthesize a stabilizing controller $K$ that minimizes the worst-case gain from exogenous inputs $\mathbf{w}$ (e.g., disturbances, sensor noises) to regulated outputs $\mathbf{z}$ (e.g., TM–SC relative motion, attitude error). This is formalized as the $H_\infty$ norm minimization problem \cite{gu2005robust}
\begin{equation}
\min_{K \in \mathcal{K}} \left\| F_\ell(P, K) \right\|_\infty,
\label{eq:H_infty_optimization}
\end{equation}
where $\mathcal{K}$ is the set of all internally stabilizing controllers, and $F_\ell(P, K)$ denotes the lower linear fractional transformation (LFT) mapping $\mathbf{w} \mapsto \mathbf{z}$:
\begin{equation}
F_\ell(P, K) = P_{11} + P_{12} K (I - P_{22} K)^{-1} P_{21}.
\label{eq:lower_fractional_transformation}
\end{equation}
The generalized plant $P$ is partitioned conformably with the input–output structure as
\[
P = 
\begin{bmatrix}
P_{11} & P_{12} \\
P_{21} & P_{22}
\end{bmatrix},
\quad
\text{where} \quad
\begin{bmatrix} \mathbf{z} \\ \mathbf{u} \end{bmatrix}
=
P
\begin{bmatrix} \mathbf{w} \\ \mathbf{y} \end{bmatrix},
\]
with $\mathbf{u}$ as the control input and $\mathbf{y}$ as the measured output.
Since analytical $H_\infty$ solutions are generally intractable, a suboptimal controller is obtained by minimizing $\eta$ subject to
\begin{equation}
\| F_{\ell}(P,K) \|_\infty \leq \eta,
\label{eq:H_infty_suboptimal}
\end{equation}
which reduces to a convex linear matrix inequality  feasibility problem via the Bounded Real Lemma \cite{LMIboyd1994linear}.

The suboptimal $H_\infty$ problem in Eq.~\eqref{eq:H_infty_suboptimal} can be solved by  Mixed Sensitivity Design~\cite{ROBUST_CONTROL}. Its inputs are the plant $G$ and weighting functions $W_1$, $W_2$, $W_3$, representing performance, control effort, and robustness requirements, respectively. The synthesis ensures the frequency-dependent bounds
\begin{align}
\| S \|_\infty &\leq \eta |W_1|^{-1}, \nonumber \\
\| KS \|_\infty &\leq \eta |W_2|^{-1}, \label{eq:controller_power_bound} \\
\| T \|_\infty &\leq \eta |W_3|^{-1}. \nonumber
\end{align}
Where the sensitivity and complementary sensitivity functions are defined as
\(
S = (I + G K)^{-1}\) and \(T = G K (I + G K)^{-1}\).
Enforcing an upper bound on $\|S\|$ improves low-frequency disturbance rejection and reference tracking; bounding $\|T\|$ ensures high-frequency robustness against sensor noise and unmodeled dynamics; and limiting $\|K S\|$ constrains control effort.Since $H_\infty$ synthesis via Mixed
Sensitivity Design requires a state-space realization, the nonlinear DFACS dynamics, i.e, Eq.~\eqref{eq:TM_SC_relative_acceleration}, Eq.~\eqref{eq:SC_angular_acceleration_CRF}, 
and Eq.~\eqref{eq:TM_attitude_dynamics} are linearized about the nominal drag-free trajectory. 
The resulting linear time-invariant model is
\begin{equation}
\dot{\mathbf{x}} = 
\begin{bmatrix}
\dot{\mathbf{p}} \\ \ddot{\mathbf{p}}
\end{bmatrix}
=
\begin{bmatrix}
\mathbf{0} & \mathbf{A}_k \\
\mathbf{A}_v & \mathbf{0}
\end{bmatrix}
\mathbf{x}
+
\begin{bmatrix}
\mathbf{0} \\ \mathbf{B}_v
\end{bmatrix}
\mathbf{u}.
\label{eq:state_space_model}
\end{equation}
Where $\mathbf{x} = [\mathbf{p}^\top\ \mathbf{\dot{p}}^\top]^\top$ is the state vector comprising relative position $\mathbf{p}$ and velocity $\mathbf{\dot{p}}$. The state variables of position are defined as
\begin{equation}
\mathbf{p} = [\mathbf{q}_{SC} \ 
\mathbf{r}_{m_{1}O_1}^{O_1}\ 
\mathbf{q}_{m_{1}O_1}  \ 
\mathbf{r}_{m_{2}O_2}^{O_2} \ 
\mathbf{q}_{m_{2}O_2}]
\in \mathbb{R}^{15},
\label{eq:state_variables}
\end{equation}
with the inputs
\begin{equation}
\mathbf{u} = 
\begin{bmatrix}
\mathbf{F}_T\ 
\mathbf{M}_T\ 
\mathbf{F}_{E_1}\ 
\mathbf{M}_{E_1}\ 
\mathbf{F}_{E_2}\ 
\mathbf{M}_{E_2}\ 
\end{bmatrix}
\in \mathbb{R}^{16},
\label{eq:input_variables_reduced}
\end{equation}
where $\mathbf{F}_{E_i} \in \mathbb{R}^2$ denotes the 2-D electrostatic force on the $i$-th TM in the plane orthogonal to its sensitive axis. $A_k$ and $A_v$ in Eq.\eqref{eq:state_space_model} is given by

\[
A_k = 
\begin{bmatrix}
I/2 & 0_{3\times3} & 0_{3\times3} & 0_{3\times3} & 0_{3\times3} \\
0_{3\times3} & I & 0_{3\times3} & 0_{3\times3} & 0_{3\times3} \\
0_{3\times3} & 0_{3\times3} & I/2 & 0_{3\times3} & 0_{3\times3} \\
0_{3\times3} & 0_{3\times3} & 0_{3\times3} & I & 0_{3\times3} \\
0_{3\times3} & 0_{3\times3} & 0_{3\times3} & 0_{3\times3} & I/2 \\
\end{bmatrix}
\in \mathbb{R}^{15\times15},
\]

\[
A_v = 
\begin{bmatrix}
0_{3\times3} & 0_{3\times3} & 0_{3\times3} & 0_{3\times3} & 0_{3\times3} \\
0_{3\times3} & S_{TT} & S_{TR} & 0_{3\times3} & 0_{3\times3} \\
0_{3\times3} & S_{RT} & S_{RR} & 0_{3\times3} & 0_{3\times3} \\
0_{3\times3} & 0_{3\times3} & 0_{3\times3} & S_{TT} & S_{TR} \\
0_{3\times3} & 0_{3\times3} & 0_{3\times3} & S_{RT} & S_{RR} \\
\end{bmatrix}
\in \mathbb{R}^{15\times15}.
\]
Where $\mathbf{S}_{TT}$, $\mathbf{S}_{TR}$, $\mathbf{S}_{RT}$ and $\mathbf{S}_{RR}$ are the stiffness matrices~\cite{diaz2013design}.

After linearization, the multiple input multiple output (MIMO) system is decoupled into 15 independent single input single output (SISO) channels via static state feedback and input allocations. Specifically, we define a desired virtual input vector $\mathbf{u}_d \in \mathbb{R}^{15}$, and compute the physical control input $\mathbf{u} \in \mathbb{R}^{16}$ as
\begin{equation}
\mathbf{u} = \mathbf{B}_v^\dagger \left( \mathbf{u}_d - \mathbf{A}_v \mathbf{p} \right),
\label{eq:decoupled_input}
\end{equation}
where $\mathbf{B}_v^\dagger \in \mathbb{R}^{16 \times 15}$ denotes the pseudoinverse of the input matrix $\mathbf{B}_v \in \mathbb{R}^{15 \times 16}$. Substituting Eq.\eqref{eq:decoupled_input} into the velocity dynamics $\ddot{\mathbf{p}} = \mathbf{A}_v \mathbf{p} + \mathbf{B}_v \mathbf{u}$ yields the fully decoupled closed-loop form
\begin{equation}
\ddot{\mathbf{p}} = \mathbf{u}_d.
\label{eq:decoupled_dynamics}
\end{equation}
Provided that \( \mathbf{B}_v \) has full row rank, the original MIMO plant is transformed into 15 uncoupled integrators, one for each relative degree of freedom, which allows for independent SISO \( H_\infty \) controller design for each channel. However, since $\mathbf{B}_v \in \mathbb{R}^{15 \times 16}$ does not have a true inverse, its generalized inverse must be computed by solving the optimization problem
\begin{equation}
\arg\min_{\mathbf{B}_v^\dagger} \| \mathbf{I}_{15} - \mathbf{B}_v \mathbf{B}_v^\dagger \|_2.
\label{eq:generalized_inverse_optimization}
\end{equation}
Here, $\mathbf{I}_{15}$ denotes the 15×15 identity matrix. Since the core task of the DFACS is to control the SC to simultaneously track the two non-orthogonal sensitive axes of the TMs, a specific control strategy for MPS is required. Denoting \[
\mathbf{r}_{m_i O_i}^{O_i} = \begin{bmatrix} x_{m_i, O_i}^{O_i}, & y_{m_i O_i}^{O_i}, & z_{m_i O_i}^{O_i} \end{bmatrix}^T.
\] 
To satisfy the drag-free control requirements, MPS is commanded to allocate its thrust in the SRF as follows: the $x$-axis thrust component generates the control force associated with the symmetric mode $x_{m_1 O_1}^{O_1} + x_{m_2 O_2}^{O_2}$, while the $y$-axis thrust component drives the antisymmetric mode $x_{m_1 O_1}^{O_1} - x_{m_2 O_2}^{O_2}$. Additionally, the $z$-axis thrust provides the control force corresponding to the transverse symmetric mode $z_{m_1 O_1}^{O_1} + z_{m_2 O_2}^{O_2}$.:
\begin{align}
\mathbf{F}_{T,x} &= B_{v,{1,4}}^\dagger \bigl( u_{d,4} + u_{d,10} \bigr)= B_{v,1,10}^\dagger \bigl( u_{d,4} + u_{d,10} \bigr).
\label{eq:x_axis_thrust} \\
\mathbf{F}_{T,y} &= B_{v,2,4}^\dagger \bigl( u_{d,4} - u_{d,10} \bigr) = B_{v,2,10}^\dagger \bigl( -u_{d,4} + u_{d,10} \bigr).
\label{eq:y_axis_thrust} \\
\mathbf{F}_{T,z} &= B_{v,3,6}^\dagger \bigl( u_{d,6} + u_{d,12} \bigr) = B_{v,3,12}^\dagger \bigl( u_{d,6} + u_{d,12} \bigr).
\label{eq:z_axis_thrust}
\end{align}

Clearly, this allocation strategy imposes constraints on Eq.~\eqref{eq:generalized_inverse_optimization}, analogous constraints for attitude channels are based on the SC and TM inertia tensors (e.g., torque allocation proportional to $\mathbf{J}_S^{-1}$, $\mathbf{J}_m^{-1}$). This convex quadratic program is solved using active-set algorithm.

After completing the decoupling, the design of the \( H_\infty \) controller continues. Eq.~\eqref{eq:decoupled_input} and Eq.~\eqref{eq:decoupled_dynamics} enable us to shift from designing a controller for \( \mathbf{u} \) to that for the decoupled SISO system \( \mathbf{u}_d \). For each degree of freedom in \( \mathbf{p} \), only the associated control requirements and noise characteristics pertinent to that specific degree of freedom need to be considered in determining the weighting functions in Eq.~\eqref{eq:controller_power_bound}. The detailed DFACS control requirements are given in Table~\ref{tab:state_precision_requirements}. 

\begin{table}[htbp]
\caption{\label{tab:state_precision_requirements}
Precision requirements for drag-free and attitude control state variables.}
\centering
\begin{tabular}{lc}
\toprule
State variable & Requirement \\
\midrule
$x_{m_iO_i}^{O_i}$ & $2.5 \times 10^{-9}\,\mathrm{m}/\sqrt{\mathrm{Hz}}$ \\
$y_{m_iO_i}^{O_i},\, z_{m_iO_i}^{O_i}$ & $1 \times 10^{-8}\,\mathrm{m}/\sqrt{\mathrm{Hz}}$ \\
$\varphi_{\mathrm{SC}},\, \theta_{\mathrm{SC}},\, \psi_{\mathrm{SC}}$ & $1 \times 10^{-8}\,\mathrm{rad}/\sqrt{\mathrm{Hz}} \times f_s$ \\
$\varphi_{m_iO_i}$ & $2 \times 10^{-7}\,\mathrm{rad}/\sqrt{\mathrm{Hz}} \times f_r$ \\
$\theta_{m_iO_i},\, \psi_{m_iO_i}$ & $1 \times 10^{-8}\,\mathrm{rad}/\sqrt{\mathrm{Hz}} \times f_s$ \\
\bottomrule
\end{tabular}
\end{table}
Based on these requirements, the weighting functions \( W_1 \) and \( W_3 \) corresponding to each degree of freedom can be calculated as
\begin{equation}
W_1 = \frac{P F_u}{B},
\label{eq:W1_weight}
\end{equation}
\begin{equation}
W_3 = \frac{P F_y}{B}.
\label{eq:W3_weight}
\end{equation}
Where \( F_u \) is the transfer function determined by the exogenous disturbance model, \( F_y \) is the transfer function determined by the measurement noise model, and \( B \) is determined by the control requirements of each state variable. In the disturbance models, special attention is given to solar radiation pressure (SRP), with its transfer function given by \cite{virdis2021meteoroid}
\begin{equation}
\mathbf{F}_{SRP} = \left( k_{dSC} + H_{press} w \right) \mathbf{T}_I^S \frac{\mathbf{r}_{SI}^I}{\| \mathbf{r}_{SI}^I \|_2}.
\label{eq:SRP_transfer_function}
\end{equation}
The corresponding torque is
\begin{equation}
\mathbf{D}_{SRP} = \begin{bmatrix} 0 & -0.6 & 0 \\ 0.6 & 0 & -0.1 \\ 0 & 0.1 & 0 \end{bmatrix} \mathbf{F}_{SRP}.
\label{eq:SRP_torque}
\end{equation}

The constant $k_{d\mathrm{SC}}$ is derived from the SC's orbital distance and the effective area of its solar arrays~\cite{SRP_Georgevic1971}. Given the similarity in orbital configuration ($\mathbf{r}_{SI}^I \approx 1~\text{AU}$) and spacecraft structural layout, we adopt the value used in the LISA mission studies, namely,
\(
k_{d\mathrm{SC}} = 6.3513 \times 10^{-5}~\text{N},
\)
which is significantly larger than the magnitude of the transfer function $H_{\mathrm{press}}$ ($\sim 1 \times 10^{-10}~\text{N}$). Therefore, including SRP directly in \( F_u \) would significantly increase the requirements on \( W_1 \) in the low-frequency range, making Eq.~\eqref{eq:controller_power_bound} difficult to solve. To address this issue, we introduce a feedforward channel that leverages spacecraft orbit determination data to estimate $\hat{\mathbf{r}}_{SI}^I$. Based on Taiji's exceptional orbital stability and its periodic ground-based orbit determination architecture, a conservative 50-km $1\sigma$ position estimation error is assumed for $\hat{\mathbf{r}}_{SI}^I$~\cite{li2021orbit}. This allows for the estimation of the ultra-low-frequency components of \( \mathbf{F}_{SRP} \) and \( \mathbf{D}_{SRP} \), denoted as \( \hat{\mathbf{F}}_{SRP,C} \) and \( \hat{\mathbf{D}}_{SRP,C} \)
\begin{equation}
\hat{\mathbf{F}}_{SRP,C} = k_{dSC} \mathbf{T}_I^S \frac{\hat{\mathbf{r}}_{SI}^I}{\| \hat{\mathbf{r}}_{SI}^I \|_2},
\label{eq:SRP_constant_estimate}
\end{equation}
\begin{equation}
\hat{\mathbf{D}}_{SRP,C} = \begin{bmatrix} 0 & -0.6 & 0 \\ 0.6 & 0 & -0.1 \\ 0 & 0.1 & 0 \end{bmatrix} \hat{\mathbf{F}}_{SRP}.
\label{eq:SRP_torque_constant_estimate}
\end{equation}
Thus, the SRP contribution to the state acceleration $\ddot{\mathbf{p}}$ is compensated by
\begin{equation}
\mathbf{\hat{a}}_{\mathrm{SRP}} = 
\begin{bmatrix} 
\mathbf{J}_S^{-1} \, \mathbf{\hat{D}}_{\mathrm{SRP},C} \\ 
m_S^{-1} \, \mathbf{T}_S^{O_1} \, \mathbf{\hat{F}}_{\mathrm{SRP}} \\ 
\boldsymbol{0}_{3 \times 1} \\ 
m_S^{-1} \, \mathbf{T}_S^{O_2} \, \mathbf{\hat{F}}_{\mathrm{SRP}} \\ 
\boldsymbol{0}_{3 \times 1} 
\end{bmatrix} .
\label{eq:SRP_acceleration_estimate}
\end{equation}
This allows Eq.~\eqref{eq:decoupled_input} to be rewritten as
\begin{equation}
\mathbf{u} = \mathbf{B}_v^\dagger \left( \mathbf{u}_d - \mathbf{A}_V \, \mathbf{\hat{p}} - \mathbf{{\hat{a}}}_{\mathrm{SRP}} \right),
\label{eq:decoupled_input_with_SRP}
\end{equation}
thereby preventing \( k_{dSC} \) from being introduced into \( F_u \).

\begin{figure}[htbp]
    \centering
    \begin{subfigure}[b]{0.48\textwidth}
        \centering
        \includegraphics[width=\textwidth]{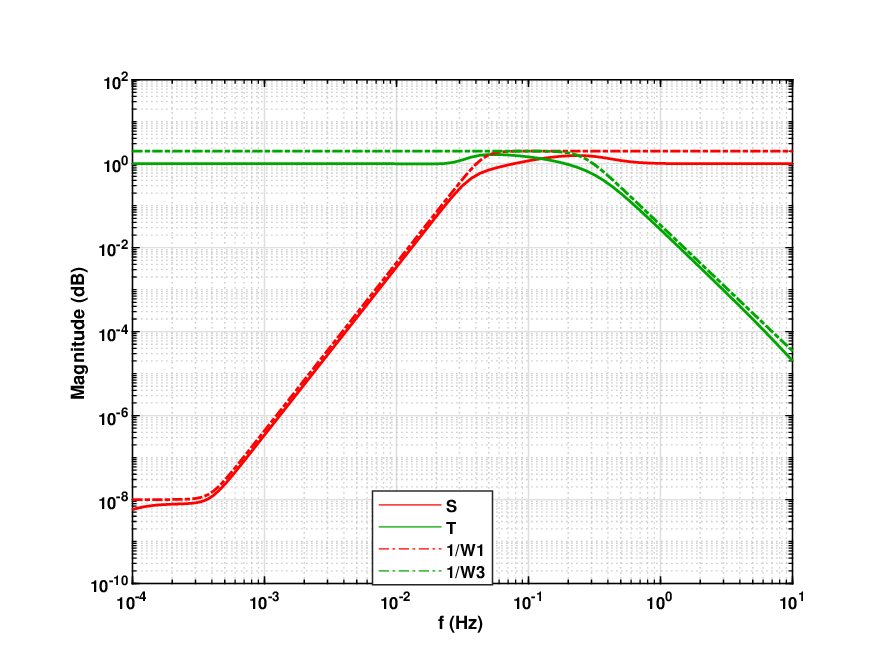}
        \caption{Results for the sensitive axis $x_{m_1 O_1}^{O_1}$ (primary performance channel), showing $S$, $T$, and their bounds $W_1^{-1}$, $W_3^{-1}$.}
    \end{subfigure}
    \hfill
    \begin{subfigure}[b]{0.48\textwidth}
        \centering
        \includegraphics[width=\textwidth]{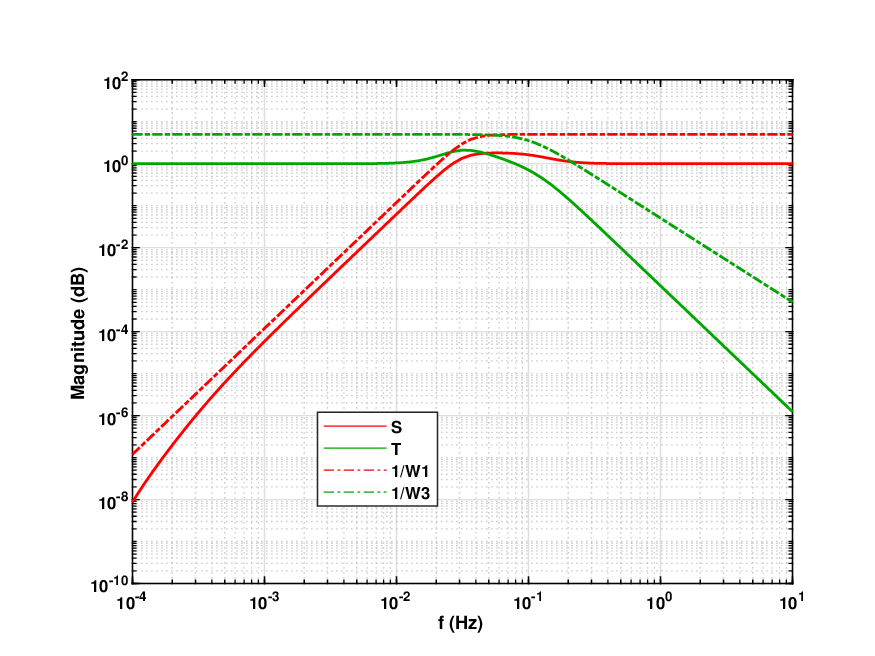}
        \caption{Corresponding $S$ and $T$ responses for the transverse axes $y_{m_1 O_1}^{O_1}$ and $z_{m_1 O_1}^{O_1}$.}
    \end{subfigure}
    \caption{Design verification of the $H_\infty$ controller: sensitivity ($S$) and complementary sensitivity ($T$) functions versus frequency. All curves remain within their prescribed bounds, demonstrating robust performance.}
    \label{fig:H_infty_results}
\end{figure}

Fig.~\ref{fig:H_infty_results} shows the mixed sensitivity $H_\infty$ controller design results. Critically, both the sensitivity function $S$ and the complementary sensitivity function $T$ satisfy their respective performance bounds over the entire bandwidth, confirming the feasibility of the robust specification.

\subsection{\label{subsec:Simulation result}Simulation result}
\begin{figure}[htbp]
\includegraphics[scale=0.53]{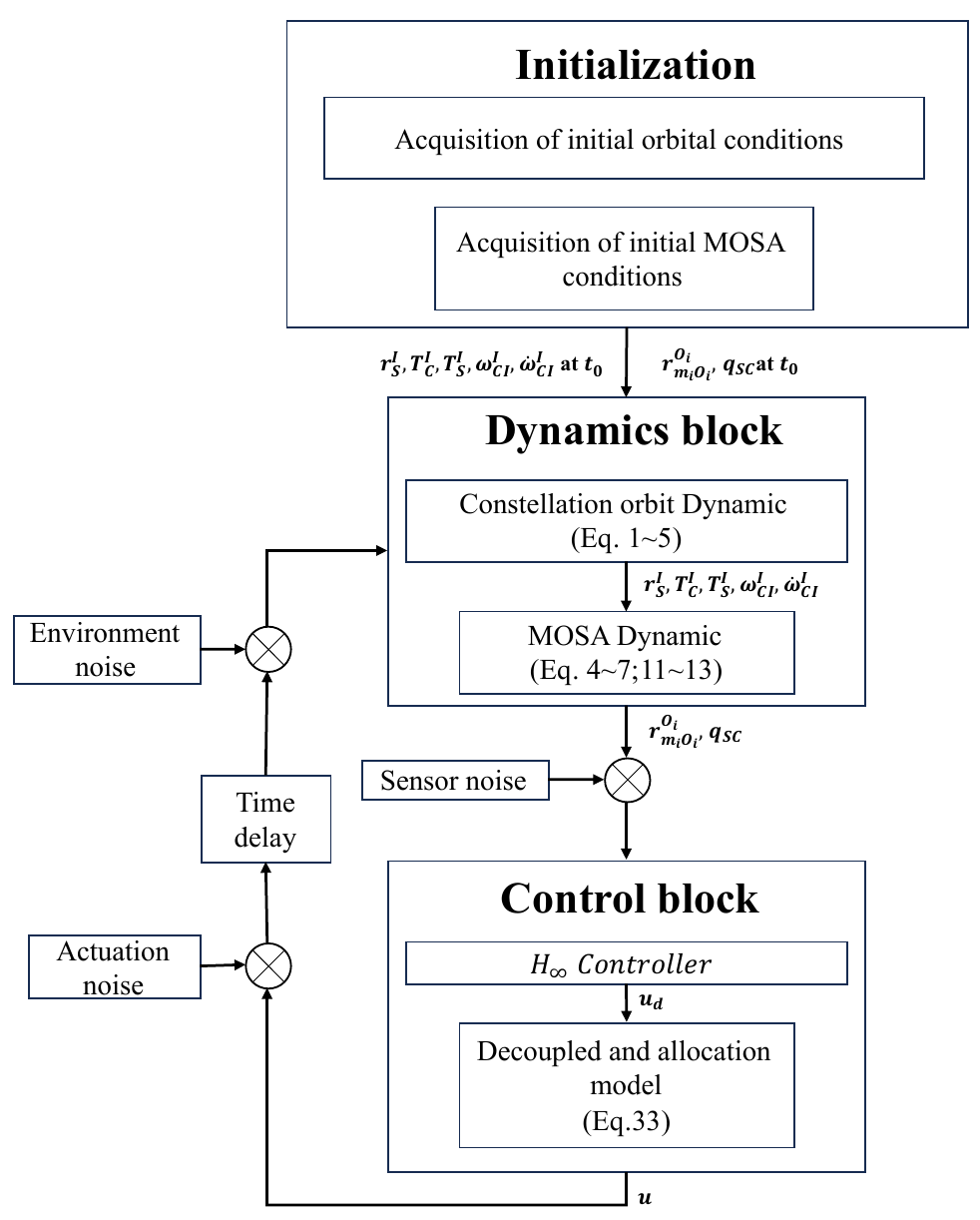}
\caption{\label{fig:flow_of_DFACS} Flow chart for the DFACS simulation.}
\end{figure}

Following the completion of the controller synthesis, a high-fidelity DFACS simulation was developed for closed-loop validation, with the architecture and data flow shown in Fig.~\ref{fig:flow_of_DFACS}.

The simulation runs at 10~Hz over 25~days, initialized at 2035-01-01T00:00:00~UTC using the planetary and lunar ephemerides DE440. Initial state perturbations of $\mathbf{p}$ are zero-mean Gaussian with $\sigma = 5 \times 10^{-8}\,\mathrm{m}$ for position variables and $5 \times 10^{-8}\,\mathrm{rad}$  for attitude variables. 

\begin{figure}[htbp]
    \centering
    \begin{subfigure}[b]{0.49\textwidth}
        \centering
        \includegraphics[width=\textwidth]{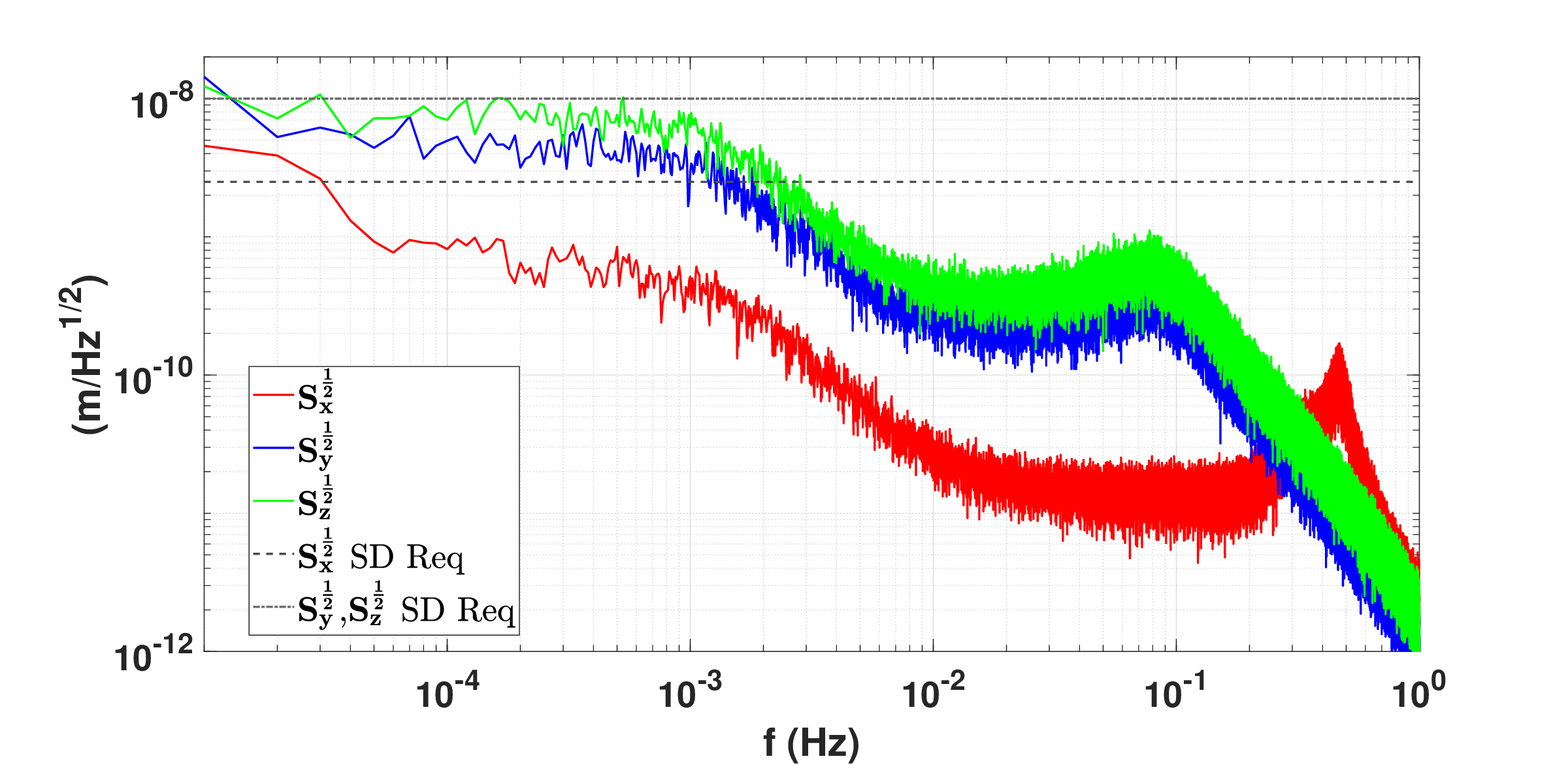}
    \end{subfigure}
    \begin{subfigure}[b]{0.49\textwidth}
        \centering
        \includegraphics[width=\textwidth]{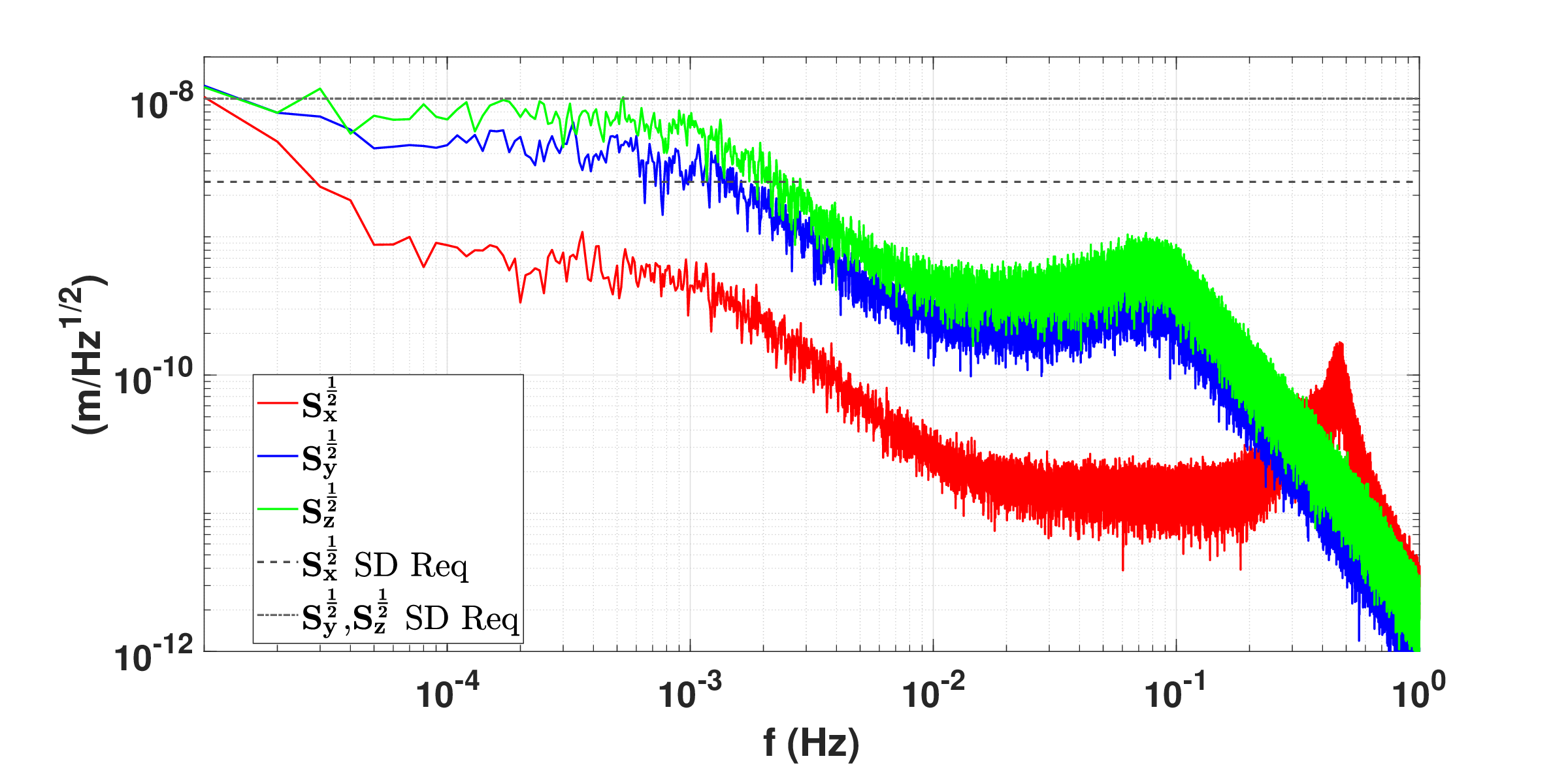}
    \end{subfigure}
    \begin{subfigure}[b]{0.49\textwidth}
        \centering
        \includegraphics[width=\textwidth]{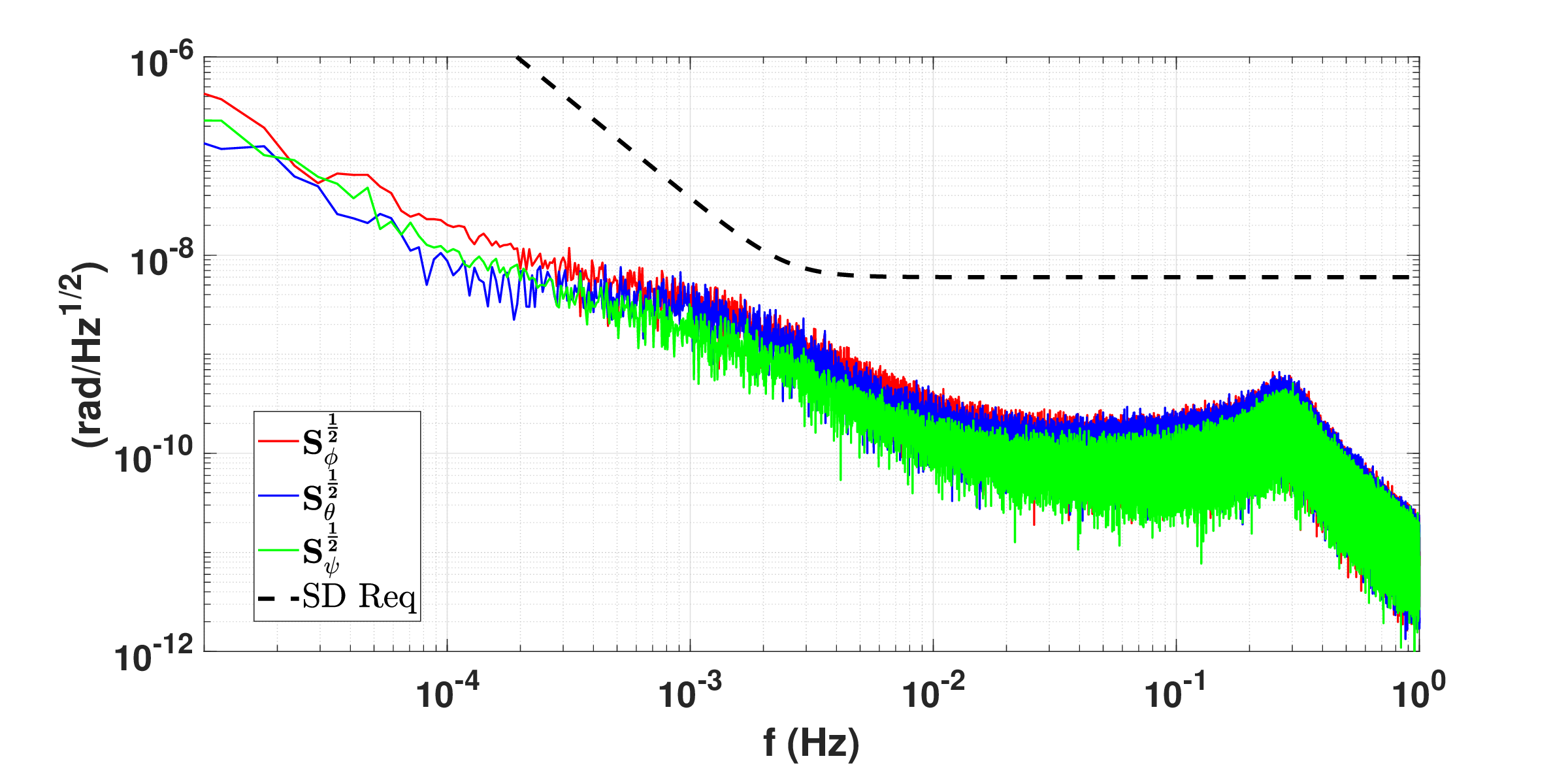}
    \end{subfigure}
    \caption{Simulation results of $\mathbf{r}_{m_iO_i}^{O_i}$ and \( \mathbf{q}_{SC} \) (converted to Euler angles). Control requirements are overlaid for reference: in the  $\mathbf{r}_{m_iO_i}^{O_i}$ subplot, the red solid line denotes the requirement along the sensitive axis, while the blue solid lines correspond to the non-sensitive axes.}
    \label{fig:r_mioi_psd}
\end{figure}

The DFACS simulation outputs both the true state $\mathbf{p}$ and its sensor readout $\tilde{\mathbf{p}}$. To assess compliance with the requirements in Table~\ref{tab:state_precision_requirements}, the amplitude spectral densities (ASDs) of $\mathbf{r}_{m_1 O_1}^{O_1}$, $\mathbf{r}_{m_2 O_2}^{O_2}$, and $\mathbf{q}_{SC}$ are plotted in Fig.~\ref{fig:r_mioi_psd}. 

\begin{table}[htbp]
\centering
\caption{The Parameters of the simulation system}
\resizebox{0.5\textwidth}{!}{
\renewcommand\cellalign{cc} 
\renewcommand{\arraystretch}{1}
\begin{tabular}{|c|c|c|}
\hline
\textbf{Parts} & \textbf{Parameters} & \textbf{Values} \\
\hline
\multirow{2}{*}{Spacecraft} 
& Mass & 1500 kg \\
\cline{2-3}
& Inertias & 
$\displaystyle J_S =
\begin{bmatrix}
800 & 13   & 10 \\
13   & 800 & 12 \\
10   & 12   & 1000
\end{bmatrix} \ \mathrm{kg \cdot m^2}$ \\
\hline
\multirow{4}{*}{TM} 
& Mass & 1.96 kg \\
\cline{2-3}
& Inertias & 
$\displaystyle J_{m} =
\begin{bmatrix}
6.913 & 0     & 0 \\
0     & 6.913 & 0 \\
0     & 0     & 6.913
\end{bmatrix} \times 10^{-4} \ \mathrm{kg \cdot m^2}$ \\
\cline{2-3}
& \makecell[c]{\vspace{2pt} SC-TM CoM offset \vspace{2pt}} & 
\makecell[c]{\vspace{2pt} $\mathbf{b}^{S}_{1} = [0.001,\ 0.18,\ 0] \ \mathrm{m}$ \\ $\mathbf{b}^{S}_{2} = [0.001,\ -0.18,\ 0] \ \mathrm{m}$ \vspace{2pt}} \\
\hline
/ & Simulation time & 25 days \\
\hline
\end{tabular}
} \label{tab_para}
\end{table}

The above simulation results demonstrate that, within the measurement frequency band of 0.1 mHz to 1 Hz, the DFACS achieves the control requirements for each degree of freedom of \( \mathbf{p} \) as specified in Table~\ref{tab:state_precision_requirements}. 


\section{\label{sec4}CoM offset calibration principle and results}
\subsection{\label{subsec4_1}Principle of CoM Offset Calibration in Science Mode}
\begin{figure}[htbp]
    \centering
    \begin{subfigure}[b]{0.52\textwidth}
        \centering
        \includegraphics[width=\textwidth]{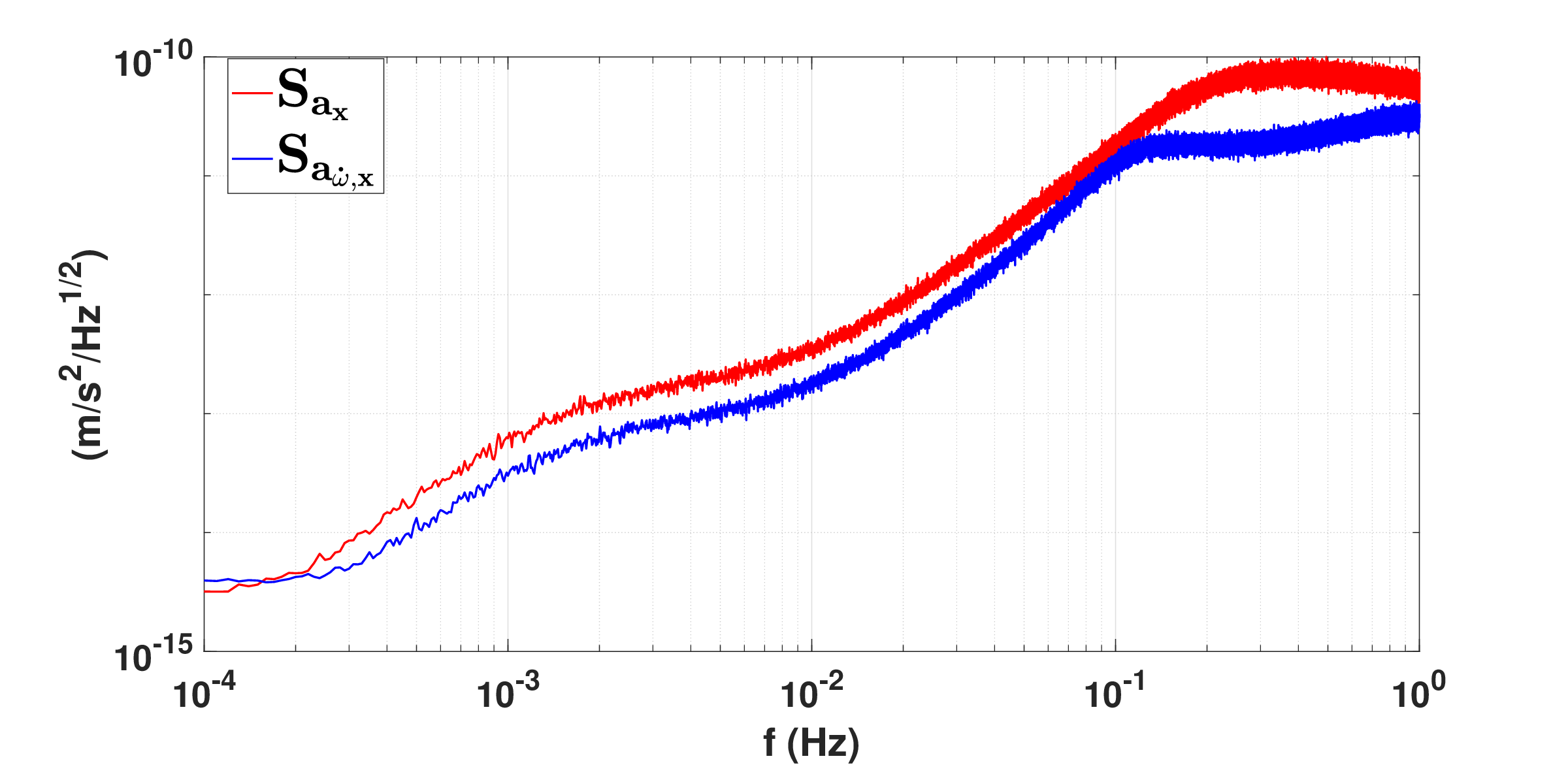}
    \end{subfigure}
    \begin{subfigure}[b]{0.52\textwidth}
        \centering
        \includegraphics[width=\textwidth]{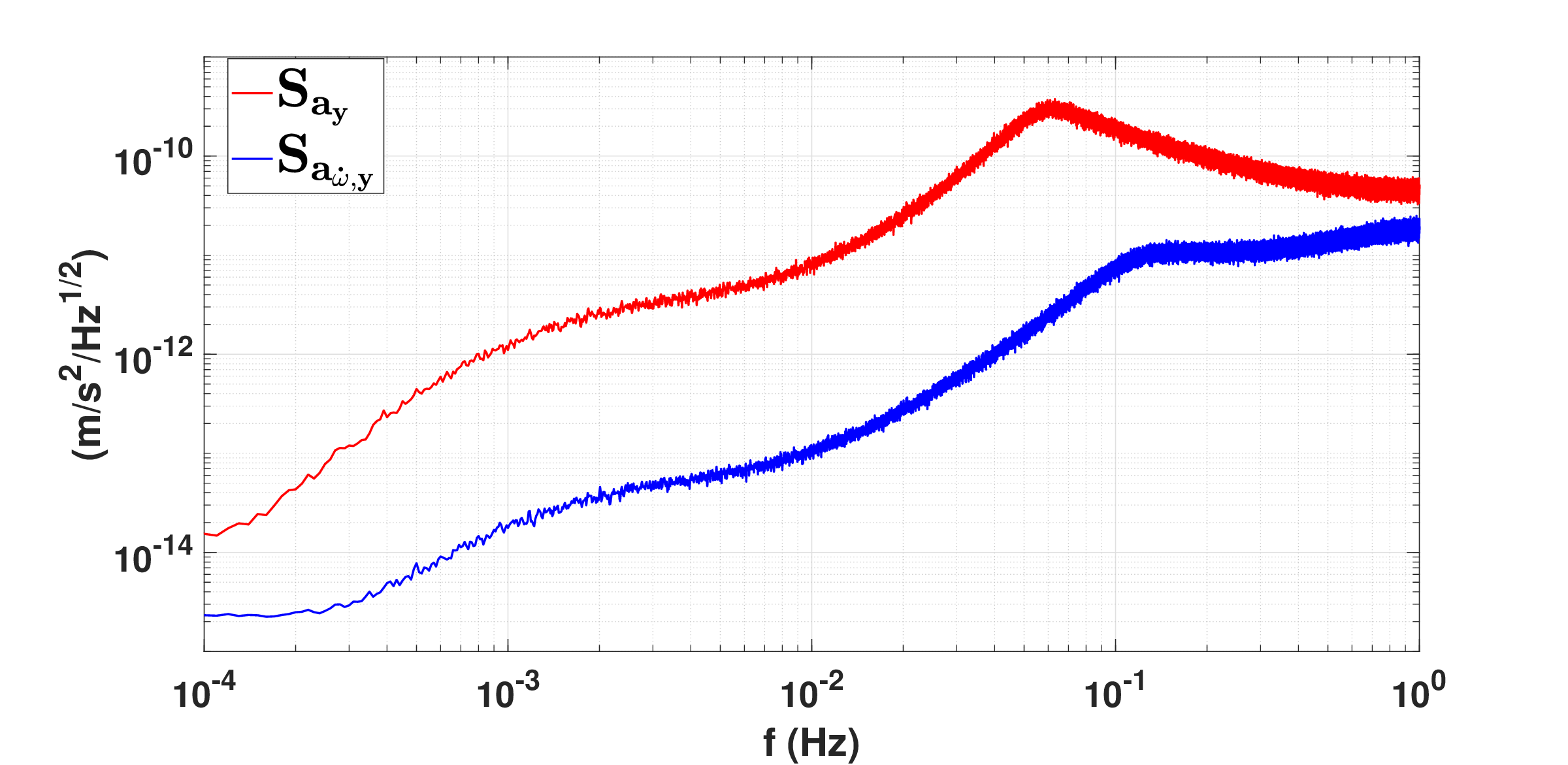}
    \end{subfigure}
    \begin{subfigure}[b]{0.52\textwidth}
        \centering
        \includegraphics[width=\textwidth]{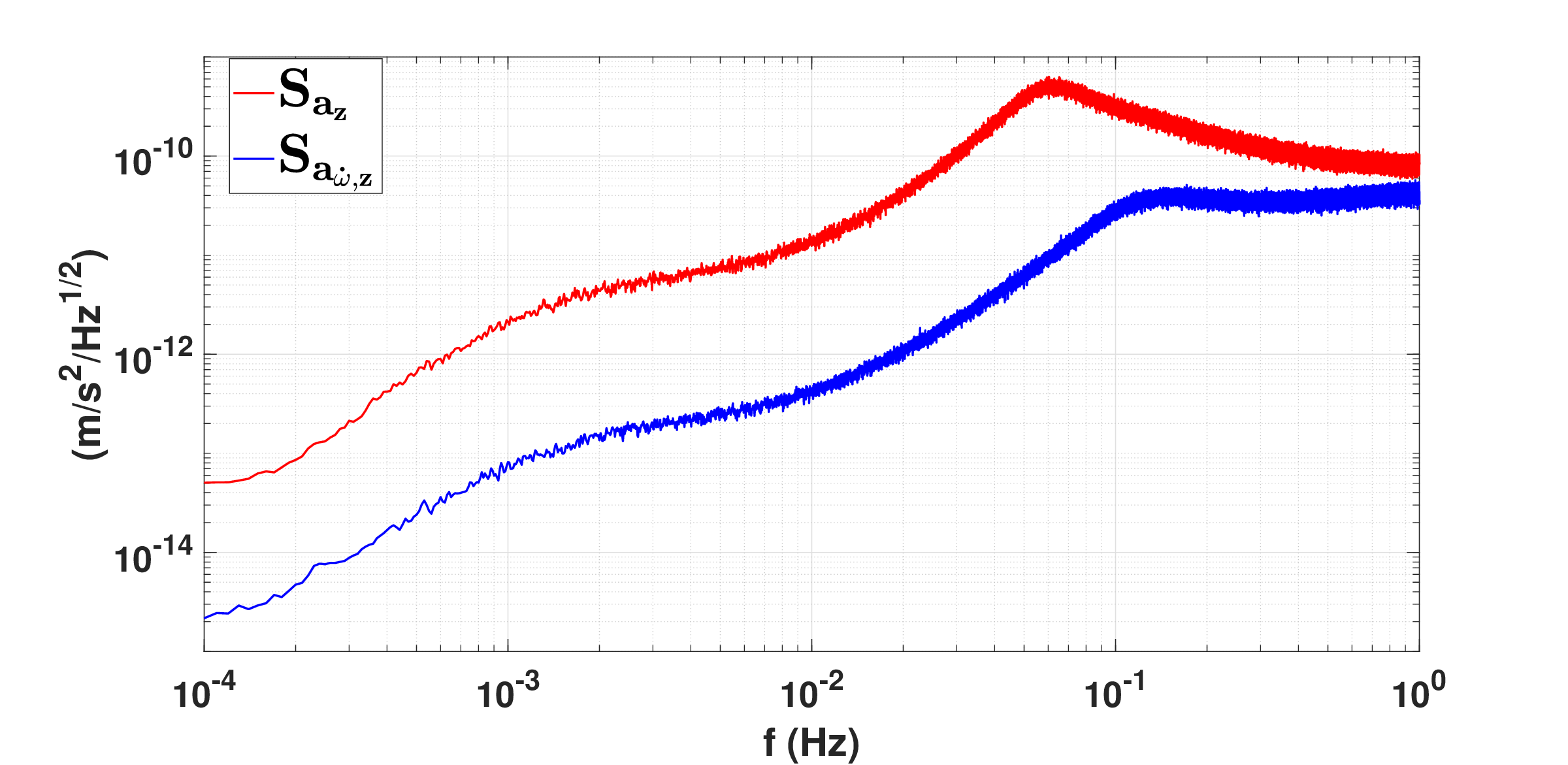}
    \end{subfigure}
    \caption{The amplitude spectral density plot of \( \mathbf{\ddot{r}}_{m_1 O_1}^{O_1} \) and \( \mathbf{a}_{\dot{\omega},1} \), from top to bottom, corresponds to the x, y, and z-axis data, with the red line representing \( \mathbf{\ddot{r}}_{m_1 O_1}^{O_1} \) and the blue line representing \( \mathbf{a}_{\dot{\omega},1} \).}
    \label{fig:acceleration_spectral_density_1}
\end{figure}

We further discuss the relationship between the acceleration \( \mathbf{\ddot{r}}_{m_i O_i}^{O_i} \) and the coupling acceleration 
\[
\mathbf{a}_{\dot{\boldsymbol{\omega}},i} \coloneqq -\boldsymbol{\Omega}\big( \mathbf{\omega}_{\mathrm{SI}}^S \big) \, \mathbf{T}_S^I \, \mathbf{b}_i^S
\]
induced by the SC attitude jitter. Based on the simulation results, the amplitude spectral density of the acceleration data \( \mathbf{\ddot{r}}_{m_1 O_1}^{O_1} \) and \( \mathbf{a}_{\dot{\boldsymbol{\omega}},1} \) is shown in Figs.~\ref{fig:acceleration_spectral_density_1}.
In the maneuver-based calibration scheme, high-amplitude square-wave angular maneuvers render the angular acceleration–induced inertial term
$\mathbf{a}_{\dot{\boldsymbol{\omega}},i}$ the dominant contribution to the TM relative acceleration $\ddot{\mathbf{r}}_{m_i O_i}^{O_i}$, 
producing a spectrally isolated peak at the maneuver frequency.
Consequently, the readout $\ddot{\mathbf{r}}_{m_1 O_i}^{O_i}$ can be modeled as,
\begin{equation}
\ddot{\mathbf{r}}_{m_i O_i}^{O_i} = \mathbf{a}_{\dot{\boldsymbol{\omega}},i} + \mathbf{a}_{\mathrm{wn}},
\label{eq:maneuver_model}
\end{equation}
where $\mathbf{a}_{\mathrm{wn}}$ denotes additive Gaussian white noise.

In contrast, during science mode calibration (i.e., without dedicated maneuvers), 
$\mathbf{a}_{\dot{\boldsymbol{\omega}},i}$ is neither dominant nor spectrally isolated, 
as evidenced by the broadband spectra in Fig.~\ref{fig:acceleration_spectral_density_1}. Instead, substituting Eq.~\eqref{eq:TM_acceleration_inertial} into Eq.~\eqref{eq:TM_SC_relative_acceleration} yields:
\begin{align}
\ddot{\mathbf{r}}_{m_i O_i}^{O_i} = \mathbf{T}_I^O \Big[ 
& \boldsymbol{\Delta}\mathbf{a}_{\mathrm{g},i} 
  + m_M^{-1} \mathbf{T}_O^I \big( \mathbf{F}_{\mathrm{E}} + \mathbf{F}_{\mathrm{st}} + \mathbf{d}_M \big) 
  \nonumber \\
& - m_S^{-1} \mathbf{T}_S^I \Big( \mathbf{F}_{\mathrm{T}} + \mathbf{d}_S 
       - \sum_{j=1}^{2} \mathbf{T}_{O_j}^S \mathbf{F}_{\mathrm{E}_j} \Big) 
  \nonumber \\
& + \mathbf{T}_S^I \, \mathbf{a}_{\dot{\boldsymbol{\omega}},i} 
\Big],
\label{eq:TM_SC_relative_acceleration_with_bias}
\end{align}
where
\(
\boldsymbol{\Delta}\mathbf{a}_{\mathrm{g},i}
= \mathbf{a}_{\mathrm{g}}(\mathbf{r}_{m_i I}^{I}) 
  - \mathbf{a}_{\mathrm{g}}(\mathbf{r}_{\mathrm{SC} I}^{I}) 
  + \mathbf{T}_{O}^{I}\,\mathbf{a}_{\mathrm{sg},i}
\)
denotes the differential gravitational acceleration between the SC and the $i$-th TM, 
with $\mathbf{a}_{\mathrm{sg},i}$ representing the self-gravitational bias. 
To account for the full dynamics of Eq.~\eqref{eq:TM_SC_relative_acceleration_with_bias}, 
we propose an extended state space representation:
\begin{align}
\dot{\mathbf{X}}_i \equiv 
\begin{bmatrix} 
  \dot{\mathbf{r}}_{m_i O_i}^{O_i} \\ 
  \ddot{\mathbf{r}}_{m_i O_i}^{O_i} \\ 
  \dot{\mathbf{a}}_{\mathrm{b},i}^{O_i} \\ 
  \dot{\mathbf{b}}_i^{O_i} \\ 
  \dot{\mathbf{n}}_i 
\end{bmatrix}
= 
\mathbf{A}_i
\begin{bmatrix} 
  \mathbf{r}_{m_i O_i}^{O_i} \\ 
  \dot{\mathbf{r}}_{m_i O_i}^{O_i} \\ 
  \mathbf{a}_{\mathrm{b},i}^{O_i} \\ 
  \mathbf{b}_i^{O_i} \\ 
  \mathbf{n}_i 
\end{bmatrix}
+ \mathbf{B}
\begin{bmatrix} 
  \mathbf{u}_i \\ 
  \mathbf{u}_{\mathrm{n},i} 
\end{bmatrix},
\label{eq:extended_state_space}
\end{align}
with the associated readout model
\begin{equation}
\mathbf{r}_{m_i O_i}^{O_i} 
= 
\mathbf{H}
\begin{bmatrix} 
  \mathbf{r}_{m_i O_i}^{O_i} \\ 
  \dot{\mathbf{r}}_{m_i O_i}^{O_i} \\ 
  \mathbf{a}_{\mathrm{b},i}^{O_i} \\ 
  \mathbf{b}_i^{O_i} \\ 
  \mathbf{n}_i 
\end{bmatrix}
+ \mathbf{D}_{\mathrm{n},i} \, \mathbf{u}_{\mathrm{n},i}.
\label{eq:readout_model}
\end{equation}

In Eq.~\eqref{eq:extended_state_space}, the state vector comprises the relative displacement $\mathbf{r}_{m_i O_i}^{O_i}$, the bias acceleration $\mathbf{a}_{\mathrm{b},i}^{O_i}$, the SC-TM CoM offset expressed in $\mathrm{ORF}_i$, $\mathbf{b}_i^{O_i} = \mathbf{T}_S^{O_i} \mathbf{b}_i^S$, and the state variables of the measurement noise shape, $\mathbf{n}_i$. The residual measurement noise is modeled as additive Gaussian white noise $\mathbf{u}_{\mathrm{n},i}$. The extended state-space matrices $\mathbf{A}_i$, $\mathbf{B}$, and $\mathbf{H}$ are defined as
\[
\mathbf{A}_{i,22\times22} = 
\begin{bmatrix} 
  \begin{bmatrix} 
    \mathbf{0}_{3 \times 3} & \mathbf{I}_{3 \times 3} & \mathbf{0}_{3 \times 6} \\  
    \mathbf{0}_{3 \times 6} & \mathbf{I}_{3 \times 3} & \boldsymbol{\Omega}_{i,3 \times 3} \\ 
    & \mathbf{0}_{6 \times 12} 
  \end{bmatrix} & \mathbf{0}_{12 \times 10} \\ 
  \mathbf{0}_{10 \times 12} & \mathbf{A}_{n,\,10 \times 10} 
\end{bmatrix},\vspace{0.2cm}
\]
\[
\mathbf{B}_{22 \times 6} = 
\begin{bmatrix} 
  \begin{bmatrix}  
    \mathbf{0}_{3 \times 3} \\ 
    \mathbf{I}_{3 \times 3} \\ 
    \mathbf{0}_{6 \times 3} 
  \end{bmatrix} & \mathbf{0}_{10 \times 3} \\ 
  \mathbf{0}_{12 \times 3} & \mathbf{B}_{n,\,10 \times 3} 
\end{bmatrix},
\]
\[
\mathbf{H}_{3 \times 22} = 
\begin{bmatrix} 
  \mathbf{I}_{3 \times 3} & \mathbf{0}_{3 \times 9} & \mathbf{C}_{n,\,3 \times 10} 
\end{bmatrix},
\]
\[
\boldsymbol{\Omega}_{i,\,3 \times 3} = 
\left[ \mathbf{T}_S^{O_i} \, \mathbf{\dot{\omega}}_{\mathrm{SI}} \right]^{\!\times}
.\]
For readability, the subscript \( i \) used to distinguish TM1 and TM2 is omitted when referring to the parameters \( \mathbf{A}_i \) in the subsequent text. 

The inputs ${\mathbf{u}}_i$, including control forces, torques, as well as the noise forces and torques, are estimated as
\[
\mathbf{\hat{u}_i} = 
\begin{bmatrix} 
  m_M^{-1} \, \mathbf{T}_O^I \, \big( \mathbf{\hat{F}}_{\mathrm{E}} + \mathbf{\hat{F}}_{\mathrm{st}} \big) 
  - m_S^{-1} \, \mathbf{T}_S^I \left( 
      \mathbf{\hat{F}}_{\mathrm{T}} + \mathbf{\hat{F}}_{\mathrm{SRP},C} 
      - \sum_{i=1}^{2} \mathbf{T}_{O_i}^S \, \mathbf{\hat{F}}_{\mathrm{E}_i} 
    \right) \\[1.2ex]
  \mathbf{\hat{M}}_{\mathrm{T}} + \mathbf{\hat{D}}_{\mathrm{SRP},C} 
  - \sum_{i=1}^{2} \mathbf{T}_{O_i}^S \, \mathbf{\hat{M}}_{\mathrm{E}_i}
\end{bmatrix}.
\]
In Eq.~\eqref{eq:extended_state_space}, $\mathbf{n}_i$ denotes the state vector of the shaped readout noise. We adopt the experimentally validated noise models from the LISA mission~\cite{vidano2022lisa}.
Since the high similarity between Taiji and LISA in scientific objectives, orbital environment, and payload configurations. Based on these established transfer-function models for the IS, IFO, and DWS, the corresponding state-space realizations for $\mathbf{n}_i$ can be constructed as
\begin{align}
&\mathbf{\dot{n}} = \mathbf{A}_n \, \mathbf{n} + \mathbf{B}_n \, \mathbf{u}_{\mathrm{n}}, \nonumber \\
&\mathbf{y}_{\mathrm{n}} = \mathbf{C}_n \, \mathbf{n} + \mathbf{D}_n \, \mathbf{u}_{\mathrm{n}}. \nonumber
\end{align}

With the extended state-space model Eqs.\eqref{eq:extended_state_space}–\eqref{eq:readout_model} established, 
the CoM offset $\mathbf{b}_i^{O_i}$ and other latent states can be jointly estimated from the DFACS readout 
$\tilde{\mathbf{r}}_{m_i O_i}^{O_i}$ through Kalman filtering methods. 
In the following, we introduce an adaptive Kalman filter algorithm that recursively updates the state estimates 
and covariance matrices, enabling robust calibration of $\mathbf{b}_i^{O_i}$ in the presence of model uncertainties 
and time-varying disturbances.
\subsection{\label{subsec4_2}Addaptive Kalman filter}
In previous work for maneuver-based CoM offset calibration, algorithms such as least squares estimation or Kalman filtering \cite{KF_ORI} have been used~\cite{wang2003study,COM_CALI_ZHANG,COM_CALI_WEI}.  Based on Eq.~\eqref{eq:extended_state_space}, the following Kalman filter is constructed:
\begin{equation}
\mathbf{\hat{X}}_{k+1}^- = \mathbf{A}_d \, \mathbf{\hat{X}}_k^+ + \mathbf{B}_d 
\begin{bmatrix} 
  \mathbf{u}_i \\ 
  \mathbf{u}_{\mathrm{n},i} 
\end{bmatrix},
\label{eq:kalman_prediction}
\end{equation}
\begin{equation}
\mathbf{d}_k = \mathbf{\tilde{r}}_{m_i O_i}^{O_i} - \mathbf{H} \, \mathbf{\hat{X}}_{k+1}^-,
\label{eq:kalman_residual}
\end{equation}
\begin{equation}
\mathbf{K}_k = \mathbf{P}_k \, \mathbf{H}^\top 
\left( \mathbf{H} \, \mathbf{P}_k \, \mathbf{H}^\top + \mathbf{R}_k \right)^{-1},
\label{eq:kalman_gain}
\end{equation}
\begin{equation}
\mathbf{\hat{X}}_{k+1}^+ = \mathbf{\hat{X}}_{k+1}^- + \mathbf{K}_k \, \mathbf{d}_k,
\label{eq:kalman_update}
\end{equation}
\begin{equation}
\mathbf{P}_{k+1} = \mathbf{A}_d \left( \mathbf{I} - \mathbf{K}_k \, \mathbf{H} \right) \mathbf{P}_k \, \mathbf{A}_d^\top + \mathbf{Q}_k.
\label{eq:kalman_covariance_update}
\end{equation}

In these equations, \( \mathbf{A}_d \) and \( \mathbf{B}_d \) are the discrete forms of \( \mathbf{A}\) and \( \mathbf{B} \) in Eq.~\eqref{eq:extended_state_space}. The subscript $k$ denotes the discrete time index corresponding to the $k$-th measurement update. The Kalman gain is denoted by \( \mathbf{K}_k \), and \( \mathbf{d}_k \) is the innovation vector. The state error covariance matrix is \( \mathbf{P}_k = \mathbb{E}\!\left\{ \boldsymbol{\delta}\mathbf{X}_k \, \boldsymbol{\delta}\mathbf{X}_k^\top \right\} \), where \( \boldsymbol{\delta}\mathbf{X}_k = \mathbf{X}_k - \mathbf{\hat{X}}_k \) represents the state error vector, defined as the difference between the true state and the estimated state at time step \(k\). The measurement noise covariance matrix is \( \mathbf{R}_k = \mathbb{E}\!\left\{ \mathbf{v}_k \, \mathbf{v}_k^\top \right\} \), and the process noise covariance matrix is \( \mathbf{Q}_k = \mathbb{E}\!\left\{ \mathbf{w}_k \, \mathbf{w}_k^\top \right\} \). 

In a standard Kalman filter, \( \mathbf{Q}_k \) and \( \mathbf{R}_k \) are given as constant matrices. Thus, for the high-dimensional extended state vector \( \mathbf{X}_{i,\,22 \times 1} \), the filter requires specifying \( \mathbf{\hat{Q}}_{i,\,22 \times 22} \) and \( \mathbf{\hat{R}}_{i,\,3 \times 3} \). However, the performance of the Kalman filter is highly sensitive to \( \mathbf{Q} \) and \( \mathbf{R} \) \cite{akf_ge2016performance,akf_ge2024adaptive}. Let the matrix elements of \( \mathbf{Q}_{n \times n} \) be \( q_{i,j} \) and those of \( \mathbf{R}_{p \times p} \) be \( r_{i,j} \), and define \( \gamma \):
\[
\gamma = q_{1,1} : q_{1,2} : \dots : q_{n,n} : r_{1,1} : r_{1,2} : \dots : r_{p,p}
\]
where $\gamma$ represents the model parameter ratio of $\mathbf{Q}$ and $\mathbf{R}$. Given estimates $\hat{\mathbf{Q}}$ and $\hat{\mathbf{R}}$, the filter is optimal if and only if $\hat{\gamma} = \gamma$. For the present extended state model, $\gamma$ is 493-dimensional, making direct iterative tuning of $\hat{\mathbf{Q}}$ and $\hat{\mathbf{R}}$ impractical. To address this, we employ an adaptive Kalman filter based on the Sage-Husa method~\cite{akf_sage1969algorithms,akf_akhlaghi2017adaptive}, which falls within the class of covariance matching techniques, a well-established approach for the estimation of process and measurement noise covariances~\cite{1972AKF_SAGEHUSA}.
\begin{equation}
\mathbf{Q}_k = \mathbf{P}_{k+1} - \mathbf{A}_d \, \mathbf{P}_k \, \mathbf{A}_d^\top + \mathbf{K}_k \, \mathbf{S}_k \, \mathbf{K}_k^\top,
\label{eq:sage_husa_update}
\end{equation}
where $\mathbf{S}_k = \mathbb{E}\!\left\{ \mathbf{d}_k \, \mathbf{d}_k^\top \right\}$ is the innovation covariance. Using exponential weighting, the iterative estimate of $\mathbf{Q}_k$ is given by
\begin{equation}
\hat{\mathbf{Q}}_k = \alpha \, \hat{\mathbf{Q}}_{k-1} + (1 - \alpha) \left[ \mathbf{P}_{k+1} - \mathbf{A}_d \, \mathbf{P}_k \, \mathbf{A}_d^\top + \mathbf{K}_k \, \mathbf{S}_k \, \mathbf{K}_k^\top \right],
\label{eq:sage_husa_Q_update}
\end{equation}
and similarly for $\mathbf{R}_k$:
\begin{equation}
\hat{\mathbf{R}}_k = \alpha \, \hat{\mathbf{R}}_{k-1} + (1 - \alpha) \left[ \mathbf{S}_k - \mathbf{H} \, \mathbf{P}_k \, \mathbf{H}^\top \right],
\label{eq:sage_husa_R_update}
\end{equation}
where $\alpha \in (0,1)$ is the adaptation gain. The updating procedure in Eqs.~\eqref{eq:sage_husa_Q_update}--\eqref{eq:sage_husa_R_update} constitutes the innovation-based covariance matching method. However, this approach does not guarantee the positive definiteness of $\hat{\mathbf{Q}}_k$ and $\hat{\mathbf{R}}_k$, which may lead to filter divergence in high-dimensional cases. To enforce positive definiteness, the residual-based covariance matching method is often adopted~\cite{song2020AKF_COMATCHING}:
\begin{equation}
\hat{\mathbf{Q}}_k = \alpha \, \hat{\mathbf{Q}}_{k-1} + (1 - \alpha) \, \mathbf{K}_k \, \mathbf{S}_k \, \mathbf{K}_k^\top,
\label{eq:residual_Q_update}
\end{equation}
\begin{equation}
\hat{\mathbf{R}}_k = \alpha \, \hat{\mathbf{R}}_{k-1} + (1 - \alpha) \left[ \mathbb{E}\!\left\{ \boldsymbol{\varepsilon}_k \, \boldsymbol{\varepsilon}_k^\top \right\} + \mathbf{H} \, \mathbf{P}_k \, \mathbf{H}^\top \right],
\label{eq:residual_R_update}
\end{equation}
where the residual $\boldsymbol{\varepsilon}_k$ is defined as
\begin{equation}
\boldsymbol{\varepsilon}_k = \tilde{\mathbf{r}}_{m_i O_i}^{O_i} - \mathbf{H} \, \hat{\mathbf{X}}_{k+1}^{+}.
\label{eq:residual_def}
\end{equation}
This formulation ensures positive definite covariance estimates and thus stable filtering. Nevertheless, by replacing $\mathbf{P}_{k+1} - \mathbf{A}_d \mathbf{P}_k \mathbf{A}_d^\top$ with zero in Eq.~\eqref{eq:residual_Q_update}, the residual-based method relaxes the statistical consistency between $\hat{\mathbf{Q}}_k$ and the true process noise, potentially degrading estimation accuracy for complex, high-dimensional systems.

Conventional covariance matching approaches are subject to two fundamental limitations: the innovation-based formulation does not guarantee the positive definiteness of the updated covariance, while the residual-based variant produces statistically inconsistent state estimates. To ensure the stability and convergence of the filter, we employ a monotonic-update scheme throughout the estimation process.

Let $|\cdot|$ denote the element-wise absolute value of a matrix,
and let $\min\{\cdot,\cdot\}$ represent the element-wise minimum operator.
The update equations are then given by
\begin{equation}
\hat{\mathbf{R}}_{\mathrm{iter}} = \hat{\mathbf{S}}_k - \mathbf{H} \, \mathbf{P}_k \, \mathbf{H}^\top,
\label{eq:akf_R_iter}
\end{equation}
\begin{equation}
\hat{\mathbf{R}}_k = \alpha \, \hat{\mathbf{R}}_{k-1} + (1 - \alpha) \, \min \left\{|\hat{\mathbf{R}}_{\mathrm{iter}}|,\, \hat{\mathbf{R}}_{k-1}\right\},
\label{eq:akf_R_update}
\end{equation}
\begin{equation}
\hat{\mathbf{Q}}_{\mathrm{iter}} = \hat{\mathbf{P}}_{k+1} - \mathbf{A}_d \, \mathbf{P}_k \, \mathbf{A}_d^\top + \mathbf{K}_k \, \hat{\mathbf{S}}_k \, \mathbf{K}_k^\top,
\label{eq:akf_Q_iter}
\end{equation}
\begin{equation}
\hat{\mathbf{Q}}_k = \alpha \, \hat{\mathbf{Q}}_{k-1} + (1 - \alpha) \, \min \left\{|\hat{\mathbf{Q}}_{\mathrm{iter}}|,\,  \hat{\mathbf{Q}}_{k-1}\right\}.
\label{eq:akf_Q_update}
\end{equation}
However, the enforced monotonicity may prevent convergence when initial estimates are overly conservative.

To resolve this, we retain the monotonic update for $\hat{\mathbf{Q}}_k$ (critical for process-noise stability) but relax it for $\hat{\mathbf{R}}_k$, exploiting the availability of reliable a priori sensor noise characterization. The factor $(1-\alpha)$ is thus shifted to the innovation residual computation,
\begin{equation}
\hat{\mathbf{R}}_{\mathrm{iter}} = (1 - \alpha) \left( \hat{\mathbf{S}}_k - \mathbf{H} \, \mathbf{P}_k \, \mathbf{H}^\top \right)
\label{eq:akf2_R_iter}
\end{equation}
\begin{equation}
\hat{\mathbf{R}}_k = \alpha \, \hat{\mathbf{R}}_{k-1} + \min \left\{ |\hat{\mathbf{R}}_{\mathrm{iter}}|,\, \hat{\mathbf{R}}_{k-1}\right\}.
\label{eq:akf2_R_update}
\end{equation}
The $\hat{\mathbf{Q}}_k$ update remains as in Eqs.~\eqref{eq:akf_Q_iter}–\eqref{eq:akf_Q_update}.  
We refer to this enhanced scheme as the proposed adaptive covariance matching method used in the CoM offset calibration. 

Finally, the Extended-State Kalman Filter is equipped with the proposed adaptive covariance matching algorithm, yielding the Extended-State Adaptive Kalman Filter (ES-AKF) formulated as follows:
\begin{equation}
\mathbf{\hat{X}}_{k+1}^- = \mathbf{A}_d \, \mathbf{\hat{X}}_k^+ + \mathbf{B}_d
\begin{bmatrix} 
  \mathbf{u}_i \\ 
  \mathbf{u}_{\mathrm{n},i} 
\end{bmatrix},
\label{eq:Final_AKF_X_k}
\end{equation}
\begin{equation}
\mathbf{d}_k = \mathbf{\tilde{r}}_{m_i O_i}^{O_i} - \mathbf{H} \, \mathbf{\hat{X}}_{k+1}^-,
\label{eq:Final_AKF_D_k}
\end{equation}
\begin{equation}
\hat{\mathbf{R}}_{\mathrm{iter}} = (1 - \alpha) \left( \hat{\mathbf{S}}_k - \mathbf{H} \, \mathbf{P}_k \, \mathbf{H}^\top \right),
\label{eq:Final_AKF_R_ITER}
\end{equation}
\begin{equation}
\hat{\mathbf{R}}_k = \alpha \, \hat{\mathbf{R}}_{k-1} + \min \left\{ |\hat{\mathbf{R}}_{\mathrm{iter}}|,\, \hat{\mathbf{R}}_{k-1}\right\},
\label{eq:Final_AKF_R_k}
\end{equation}
\begin{equation}
\mathbf{K}_k = \mathbf{P}_k \, \mathbf{H}^\top 
\left( \mathbf{H} \, \mathbf{P}_k \, \mathbf{H}^\top + \mathbf{R}_k \right)^{-1},
\label{eq:Final_AKF_K}
\end{equation}
\begin{equation}
\mathbf{\hat{X}}_{k+1}^+ = \mathbf{\hat{X}}_{k+1}^- + \mathbf{K}_k \, \mathbf{d}_k,
\label{eq:Final_AKF_x_k1}
\end{equation}
\begin{equation}
\hat{\mathbf{Q}}_{\mathrm{iter}} = \hat{\mathbf{P}}_{k+1} - \mathbf{A}_d \, \mathbf{P}_k \, \mathbf{A}_d^\top + \mathbf{K}_k \, \hat{\mathbf{S}}_k \, \mathbf{K}_k^\top,
\label{eq:Final_AKF_Q_ITER}
\end{equation}
\begin{equation}
\hat{\mathbf{Q}}_k = \alpha \, \hat{\mathbf{Q}}_{k-1} + (1 - \alpha) \, \min \left\{|\hat{\mathbf{Q}}_{\mathrm{iter}}|,\,  \hat{\mathbf{Q}}_{k-1}\right\},
\label{eq:Final_AKF_Q_K}
\end{equation}
\begin{equation}
\mathbf{P}_{k+1} = \mathbf{A}_d \left( \mathbf{I} - \mathbf{K}_k \, \mathbf{H} \right) \mathbf{P}_k \, \mathbf{A}_d^\top + \mathbf{Q}_k.
\label{eq:Final_AKF_p_k1}
\end{equation}


\begin{figure}[htbp]
\centering
\includegraphics[scale=0.24]{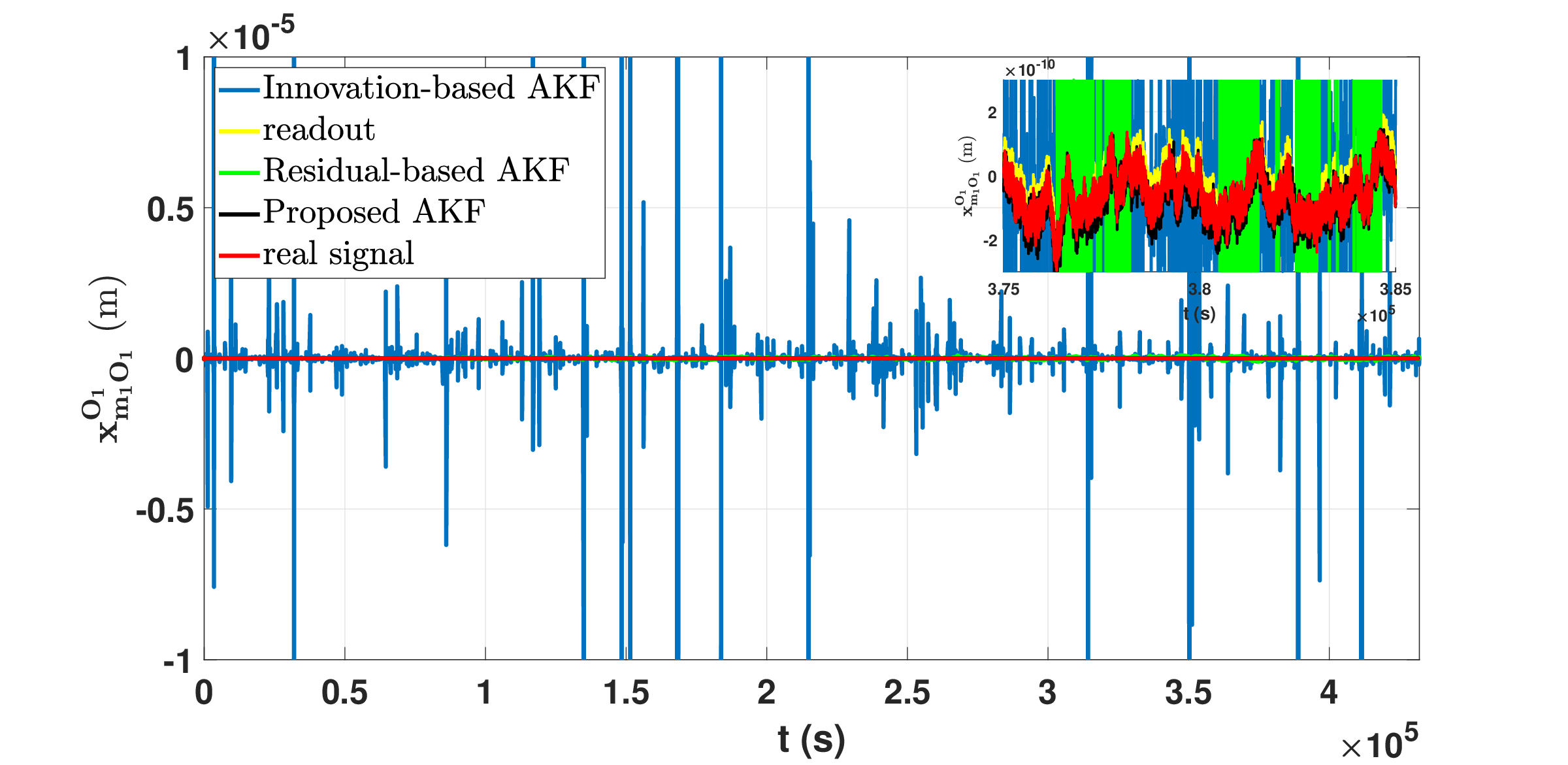}
\caption{Adaptive Kalman filter results of \( \tilde{x}_{m_1O_1}^{O_1} \).}
\label{fig:akf1_vs_akf2_x}
\includegraphics[scale=0.24]{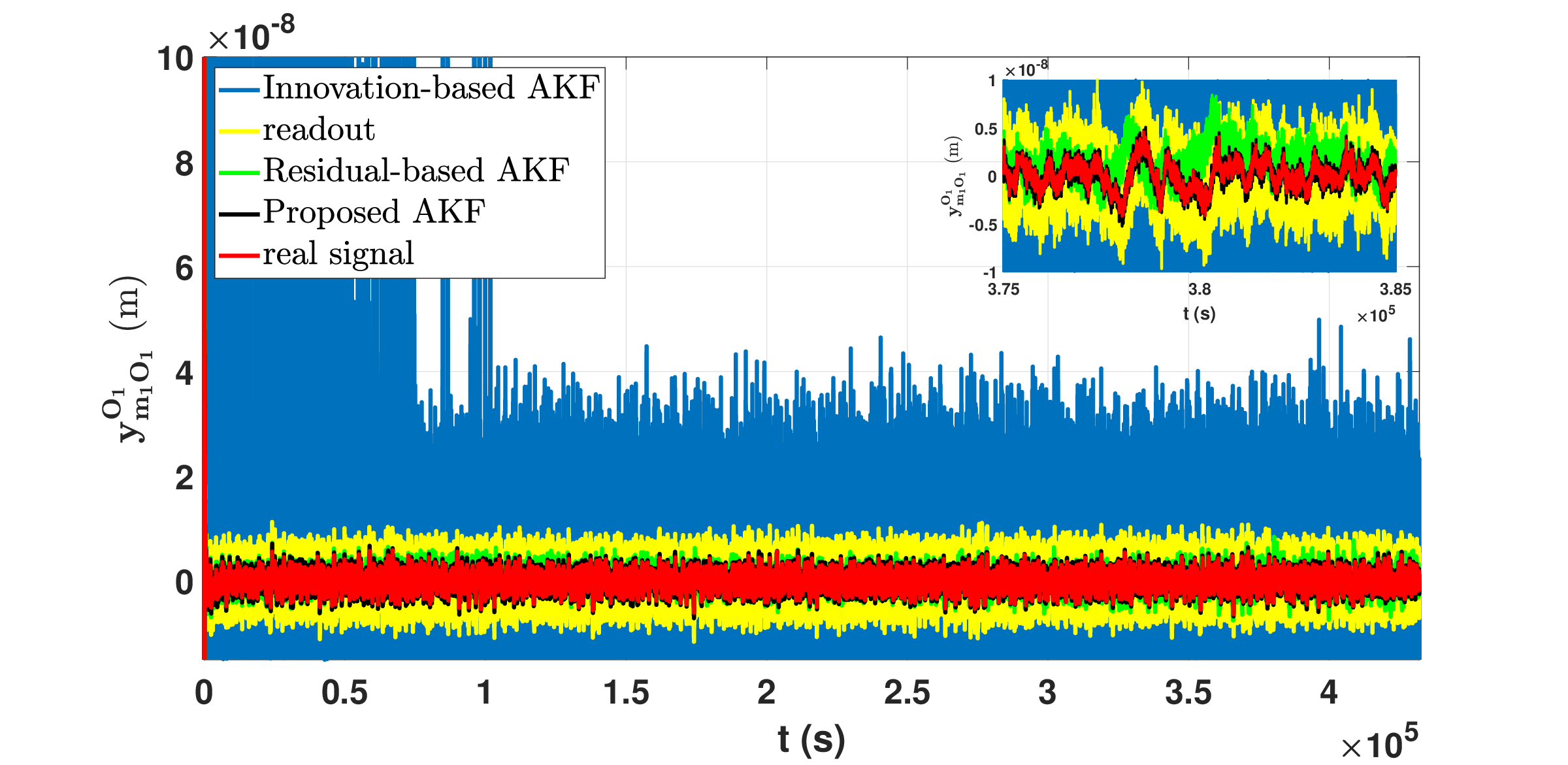}
\caption{Adaptive Kalman filter results of \( \tilde{y}_{m_1O_1}^{O_1} \).}
\label{fig:akf1_vs_akf2_y}
\includegraphics[scale=0.24]{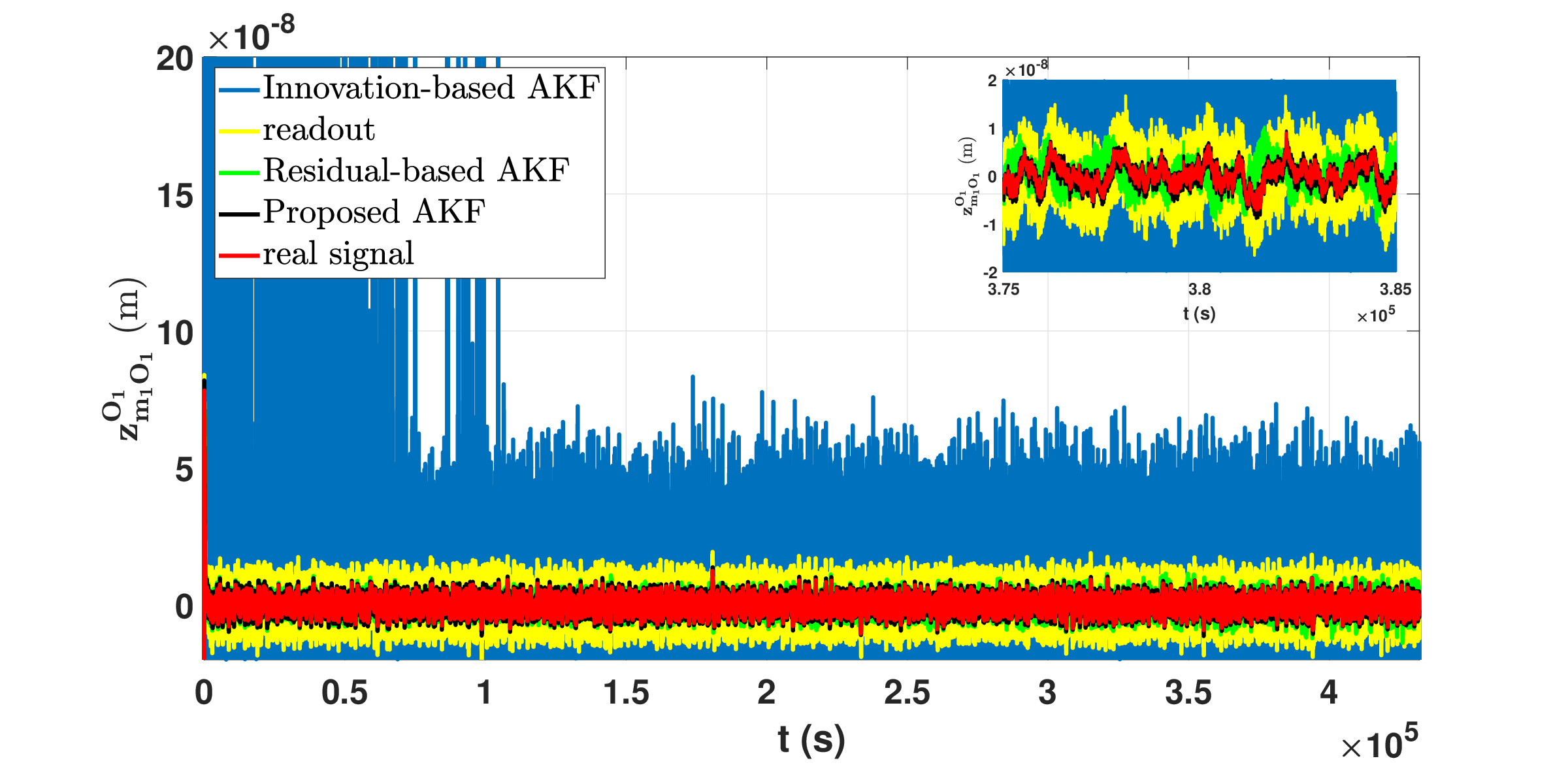}
\caption{Adaptive Kalman filter results of \( \tilde{z}_{m_1O_1}^{O_1} \).}
\label{fig:akf1_vs_akf2_z}
\end{figure}

Based on Eqs.\eqref{eq:Final_AKF_X_k}-\eqref{eq:Final_AKF_p_k1}, the extended state $\mathbf{X}_i$ is estimated using the proposed ES-AKF. For comparison, its performance is evaluated against the innovation-based and residual-based methods on the TM1–SC relative displacement readout $\tilde{\mathbf{r}}_{m_1 O_1}^{O_1}$.
Figs.~\ref{fig:akf1_vs_akf2_x}--\ref{fig:akf1_vs_akf2_z} compare the filtering performance of three adaptive Kalman filter variants: the innovation-based method, the residual-based method, and the proposed approach. These are evaluated on the TM1–SC relative displacement $\mathbf{r}_{m_1 O_1}^{O_1}$ along its three measurement axes. The plots show the true trajectory (red), the raw readout data (yellow), the innovation-based estimate (blue), the residual-based estimate (green), and the proposed estimate (black).

Along all three axes, the innovation-based AKF (blue) exhibits severe noise amplification due to instability arising from non-positive-definite covariance updates. This is most evident for the $x$-axis, where the filter generates high-amplitude spikes that exceed the measurement range. The residual-based AKF (green) avoids such instabilities but introduces a noticeable phase lag relative to the true signal, particularly in regions of rapid motion. In contrast, the proposed AKF (black) achieves near-perfect tracking of the true trajectory across all axes. It suppresses measurement noise without introducing phase distortion or amplitude bias, thereby delivering higher fidelity and precise temporal alignment with the ground truth. 

Therefore, in light of its superior noise suppression, phase fidelity, and robust convergence, we will adopt the proposed AKF for CoM offset calibration.

\subsection{\label{subsec4_3}Simulation Result}

\begin{figure}[htbp]
\centering
\includegraphics[scale=0.25]{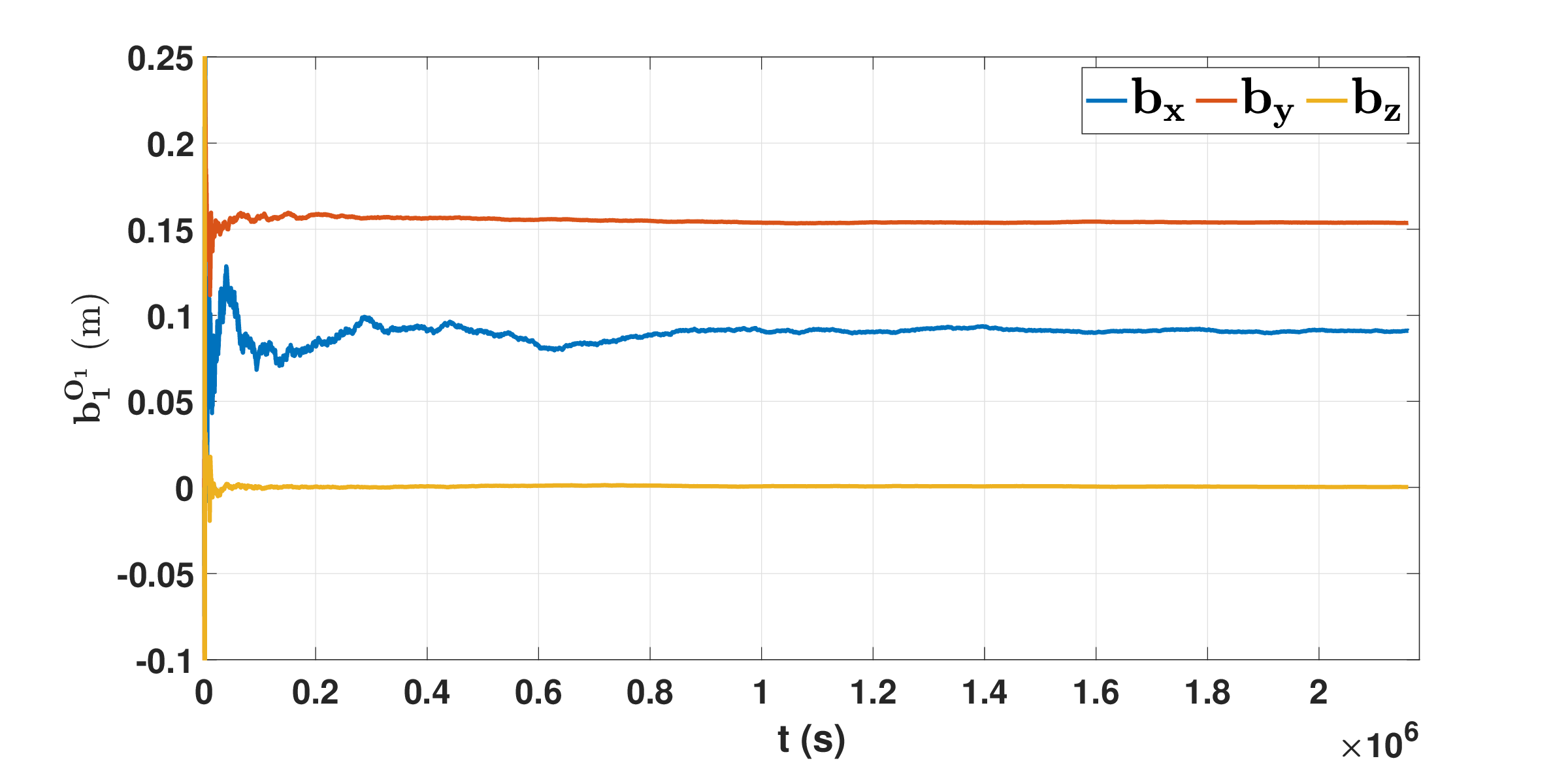}
\caption{\label{fig:tm1_com_convergence} The CoM offset calibration result \( \mathbf{b}_1^{O_1} \).}
\end{figure}

\begin{figure}[htbp]
\centering
\includegraphics[scale=0.25]{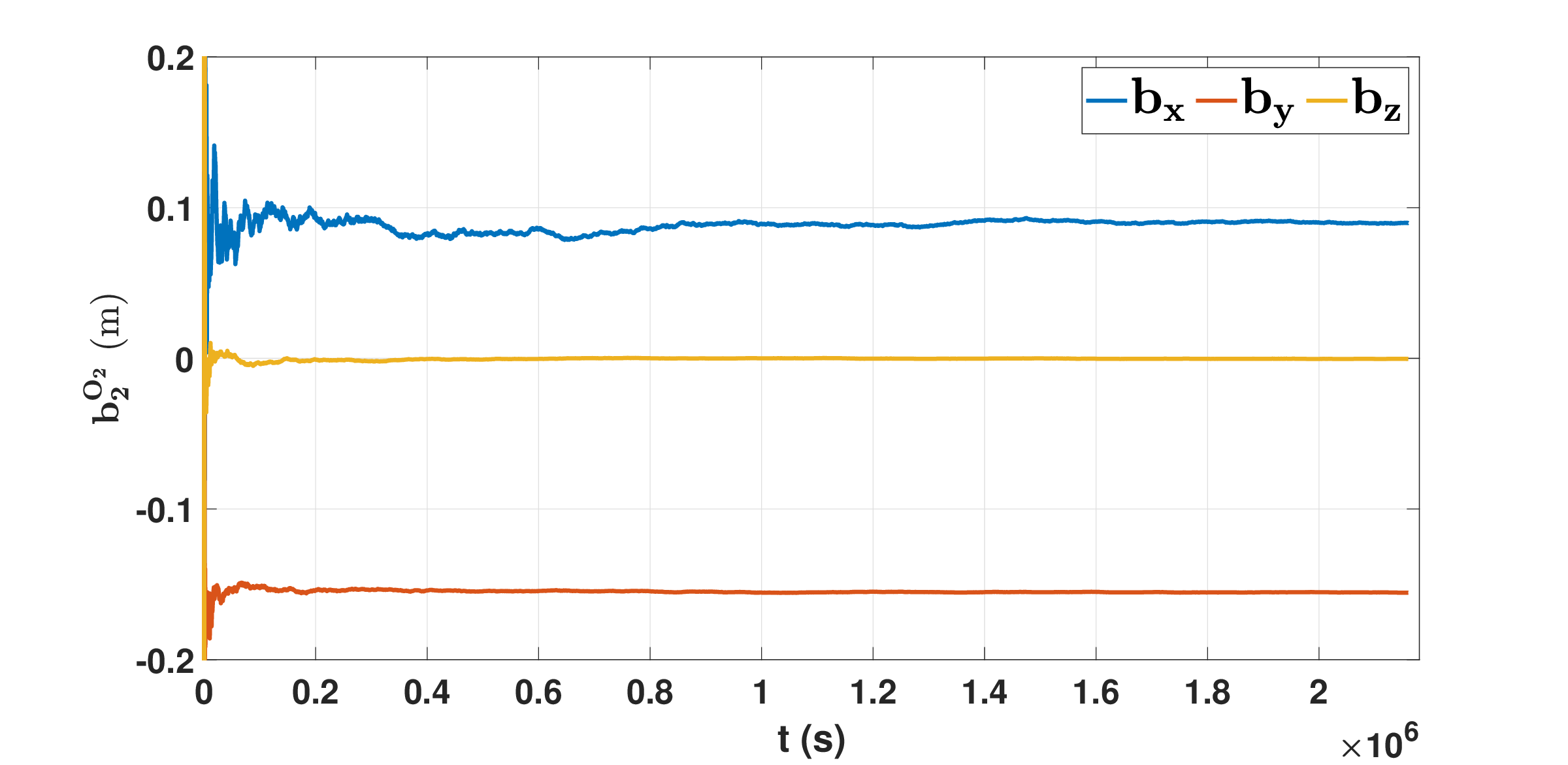}
\caption{\label{fig:tm2_com_convergence} The CoM offset calibration result \( \mathbf{b}_2^{O_2} \).}
\end{figure}

Using the science mode readouts, the ES-AKF estimates the CoM offsets $\mathbf{b}_1^{O_1}$ and $\mathbf{b}_2^{O_2}$ of TM1 and TM2 relative to the SC. The convergence behavior of these estimates is presented in Figs.~\ref{fig:tm1_com_convergence} and~\ref{fig:tm2_com_convergence}.

Figs.~\ref{fig:tm1_com_convergence} and~\ref{fig:tm2_com_convergence} show that the estimates of $\mathbf{b}_{i,x}^{O_i}$ exhibit slower convergence compared to those of $\mathbf{b}_{i,y}^{O_i}$ and $\mathbf{b}_{i,z}^{O_i}$. This behavior arises because the $x$-axis CoM calibration relies solely on IS readout signals ($\tilde{y}_{m_i O_i}^{O_i}$, $\tilde{z}_{m_i O_i}^{O_i}$), which have lower precision, whereas the $y$- and $z$-axis calibrations additionally utilize the high-precision IFO readout ($\tilde{x}_{m_i O_i}^{O_i}$). Consequently, \( \mathbf{b}_{i,y}^{O_i} \) and \( \mathbf{b}_{i,z}^{O_i} \) converge using data of approximately \( 0.4 \times 10^6~\mathrm{s} \) in duration, while the \( \mathbf{b}_{i,x}^{O_i} \) requires approximately \( 1.4 \) to \( 1.6 \times 10^6~\mathrm{s} \). Correspondingly, for the confidence level determined from the error covariance matrix, the final calibration accuracy of \( \mathbf{b}_{1,x}^{O_1} \) is lower than that of \( \mathbf{b}_{1,y}^{O_1} \) and \( \mathbf{b}_{1,z}^{O_1} \). Table~\ref{tab:com_calibration_results} lists the calibrated values, the true values, and the confidence levels computed from the final state error covariance matrix \( \mathbf{P}_{\mathrm{end}} \) (taken as \( \mathbf{P}_{\infty} \)) at the termination time.
\begin{table*}[htbp]
\caption{\label{tab:com_calibration_results}SC-TM CoM Offset Calibration Results}
\begin{ruledtabular}
\centering
\begin{tabular}{
    c
    S[table-format=-1.4e-1]
    S[table-format=1.4e-1]
    S[table-format=-1.4e-1]
}
\textbf{State variable} & \textbf{Calibration Value (Unit: m)} & \textbf{Uncertainty (Unit: m)} & \textbf{True Value (Unit: m)} \\ \hline
$\mathbf{b}_{1,x}^{O_1}$ & 9.081e-2  & 1.335e-3  & 9.087e-2  \\
$\mathbf{b}_{1,y}^{O_1}$ & 1.540e-1  & 1.705e-4  & 1.554e-1  \\
$\mathbf{b}_{1,z}^{O_1}$ & 5.086e-4  & 9.219e-5  & 0         \\
$\mathbf{b}_{2,x}^{O_2}$ & 9.019e-2  & 3.762e-3  & 9.087e-2  \\
$\mathbf{b}_{2,y}^{O_2}$ & -1.553e-1 & 1.394e-4  & -1.554e-1 \\
$\mathbf{b}_{2,z}^{O_2}$ & -3.509e-4 & 7.746e-5  & 0         \\
\end{tabular}
\end{ruledtabular}
\end{table*}

Based on the results in Table~\ref{tab:com_calibration_results}, the system identification algorithm using the ES-AKF estimator achieves high-precision calibration of the SC-TM CoM offset in science mode using only IS/IFO/DWS readout signals. The confidence levels are better than \( 4~\mathrm{mm} \) for \( \mathbf{b}_{i,x}^{O_i} \), better than \( 0.2~\mathrm{mm} \) for \( \mathbf{b}_{i,y}^{O_i} \), and better than \( 0.1~\mathrm{mm} \) for \( \mathbf{b}_{i,z}^{O_i} \). However, the calibrated value of \( \mathbf{b}_{i,x}^{O_i} \) is actually closer to the true value than that of \( \mathbf{b}_{i,y}^{O_i} \). This can be explained by the following two reasons.

First, the calibration value is determined as the mathematical expectation of all estimates over the final \( 3600~\mathrm{s} \) of data. In theory, the estimate of \( \mathbf{b}_i^{O_i} \) oscillates around or slowly approaches the true value. Therefore, the close agreement between the \( \mathbf{b}_{i,x}^{O_i} \) estimate and the true value indicates that its convergence point is near the true value. Nevertheless, its error variance—i.e., the amplitude of oscillation—remains the largest among the three degrees of freedom. This behavior is clearly illustrated by the convergence curves in Figs.~\ref{fig:tm1_com_convergence} and~\ref{fig:tm2_com_convergence}.

Second, the relative displacement \( \mathbf{r}_{m_i O_i}^{O_i} \) exhibits higher noise in the ultra-low-frequency band below the control bandwidth along the sensitive axis (as shown in Fig.~\ref{fig:r_mioi_psd}), which does not affect control requirements but significantly impacts the estimation accuracy of \( \mathbf{x}_{m_i O_i}^{O_i} \) by the AKF in the ultra-low-frequency range (see Fig.~\ref{fig:akf1_vs_akf2_x}), thereby affecting the estimation of \( \mathbf{b}_{i,y}^{O_i} \).

Overall, comparing the true values with the estimated results, the estimation errors of the SC-TM CoM offset using the ES-AKF estimator—based solely on scientific data readout—are less than \( 1\% \).

\section{\label{sec5}Conclusion}

This work proposes and validates a maneuver-free calibration scheme for the SC-TM CoM offset during the science operation of space-based GW antennas, especially for Taiji (a typical LISA-like mission). The method directly utilizes the standard science-mode readouts of the IS, interferometer, and differential wavefront sensor, eliminating the need for dedicated spacecraft maneuvers that disrupt measurement continuity and laser-link stability.

A high-fidelity, closed-loop simulation framework was developed, integrating a detailed nonlinear dynamical model of the LISA-like SC-TM system and a robust $H_{\infty}$ drag-free and attitude controller. Using 25 days of simulated science data, we designed an Extended-State Adaptive Kalman Filter that incorporates the CoM offset induced acceleration noises, sensor noise dynamics, and systematic biases into a unified state-estimation framework. The filter successfully identifies the CoM offset vector with an accuracy of 0.01–1.5 mm across all axes, corresponding to a maximum relative error below 1\%.
Beyond calibration, the proposed ES-AKF demonstrates the capability of high-precision filtering of the IS readout signal and suppression of noise. 



In summary, this study provides a viable and precise calibration strategy that can operate continuously during science observations, enabling the largest nonlinear disturbance in the DFACS to be accurately predicted. The framework also contributes a model-based estimation tool that supports advanced DFACS functions including disturbance compensation, performance diagnostics, and nonlinear control design.
\section*{Acknowledgements}
This work was supportedin partby the National Key R\&D Program of China under Grant 2021YFC2202900, in part by the National Natural Science Foundation of China under Grant 62203145 and U25B2052, in part by the Heilongjiang Provincial Natural Science Foundation under Grant YQ2024F010, and in part by the International Partnership Program of the Chinese Academy of Sciences, Grant No.
025GJHZ2023106GC.

\bibliography{main}

@article{TQ1_RESULT,
doi = {10.1088/1361-6382/aba66a},
url = {https://dx.doi.org/10.1088/1361-6382/aba66a},
year = {2020},
month = {aug},
publisher = {IOP Publishing},
volume = {37},
number = {18},
pages = {185013},
author = {Luo, Jun and Bai, Yan-Zheng and Cai, Lin and Cao, Bin and Chen, Wei-Ming and Chen, Yu and Cheng, De-Cong and Ding, Yan-Wei and Duan, Hui-Zong and Gou, Xingyu and Gu, Chao-Zheng and Gu, De-Feng and He, Zi-Qi and Hu, Shuang and Hu, Yuexin and Huang, Xiang-Qing and Jiang, Qinghua and Jiang, Yuan-Ze and Li, Hong-Gang and Li, Hong-Yin and Li, Jia and Li, Ming and Li, Zhu and Li, Zhu-Xi and Liang, Yu-Rong and Liao, Fang-Jie and Liu, Yan-Chong and Liu, Li and Liu, Pei-Bo and Liu, Xuhui and Liu, Yuan and Lu, Xiong-Fei and Luo, Yan and Mei, Jianwei and Ming, Min and Qu, Shao-Bo and Tan, Ding-Yin and Tang, Mi and Tu, Liang-Cheng and Wang, Cheng-Rui and Wang, Fengbin and Wang, Guan-Fang and Wang, Jian and Wang, Lijiao and Wang, Xudong and Wei, Ran and Wu, Shu-Chao and Xiao, Chun-Yu and Xie, Meng-Zhe and Xu, Xiao-Shi and Yang, Liang and Yang, Ming-Lin and Yang, Shan-Qing and Yeh, Hsien-Chi and Yu, Jian-Bo and Zhang, Lihua and Zhao, Meng-Hao and Zhou, Ze-Bing},
title = {The first round result from the TianQin-1 satellite},
journal = {Classical Quantum Gravity}
}

@article{LIGO2016,
  title = {Observation of Gravitational Waves from a Binary Black Hole Merger},
  author = {Abbott, B. P. and others},
  collaboration = {LIGO Scientific Collaboration and Virgo Collaboration},
  journal = {Phys. Rev. Lett.},
  volume = {116},
  issue = {6},
  pages = {061102},
  numpages = {16},
  year = {2016},
  month = {Feb},
  publisher = {American Physical Society},
  doi = {10.1103/PhysRevLett.116.061102},
  url = {https://link.aps.org/doi/10.1103/PhysRevLett.116.061102}
}

@article{TAIJI_OVERVIEW1,
author = {Ruan, Wen-Hong and Guo, Zong-Kuan and Cai, Rong-Gen and Zhang, Yuan-Zhong},
title = {Taiji program: Gravitational-wave sources},
journal = {	Int. J. Mod. Phys. A},
volume = {35},
number = {17},
pages = {2050075},
year = {2020},
doi = {10.1142/S0217751X2050075X},
URL = {https://doi.org/10.1142/S0217751X2050075X}
}

@article{TQ_OVERVIEW,
  title={TianQin: a space-borne gravitational wave detector},
  author={Luo, Jun and others},
  journal={Classical Quantum Gravity},
  volume={33},
  number={3},
  pages={035010},
  year={2016},
  publisher={IOP Publishing},
  doi = {10.1088/0264-9381/33/3/035010},
  url = {https://dx.doi.org/10.1088/0264-9381/33/3/035010},
}

@article{LPF2018calibrating,
  title={Calibrating the system dynamics of LISA Pathfinder},
  author={Armano, Michele and Audley, H and Baird, J and Binetruy, P and Born, M and Bortoluzzi, D and Castelli, E and Cavalleri, A and Cesarini, A and Cruise, AM and others},
  journal={Physical Review D},
  volume={97},
  number={12},
  pages={122002},
  year={2018},
  publisher={APS},
  doi = {10.1103/PhysRevD.97.122002}
}

@article{luo2020taiji,
  title={The Taiji program: A concise overview},
  author={Luo, Ziren and Wang, Yan and Wu, Yueliang and Hu, Wenrui and Jin, Gang},
  journal={Progress of Theoretical and Experimental Physics},
  volume={2021},
  number={5},
  pages={05A108},
  year={2020},
  publisher={Oxford University Press},
  doi = {10.1093/ptep/ptaa083},
  url = {https://doi.org/10.1093/ptep/ptaa083},
}

@article{TAIJI-1_OVERVIEW,
    author = "Cai, Zhiming and Deng, Jianfeng",
    title = "{Satellite architecture and preliminary in-orbit experiment of Taiji-1}",
    doi = "10.1142/S0217751X21400200",
    journal = "Int. J. Mod. Phys. A",
    volume = "36",
    number = "11n12",
    pages = "2140020",
    year = "2021"
}

@article{COM_CALI_WANG_JOURNAL,
  title={Determination of center-of-mass of gravity recovery and climate experiment satellites},
  author={Wang, Furun and Bettadpur, Srinivas and Save, Himanshu and Kruizinga, Gerhard},
  journal={J. Spacecr. Rockets},
  volume={47},
  number={2},
  pages={371--379},
  year={2010},
  doi={10.2514/1.46086},
  url={https://doi.org/10.2514/1.46086}
}

@Article{COM_CALI_TH,
AUTHOR = {Huang, Zhiyong and Li, Shanshan and Cai, Lin and Fan, Diao and Huang, Lingyong},
TITLE = {Estimation of the Center of Mass of GRACE-Type Gravity Satellites},
JOURNAL = {Remote Sens.},
VOLUME = {14},
YEAR = {2022},
NUMBER = {16},
ARTICLE-NUMBER = {4030},
URL = {https://doi.org/10.3390/rs14164030},
ISSN = {2072-4292}
}

@phdthesis{wang2003study,
  title={Study on center of mass calibration and K-band ranging system calibration of the GRACE mission},
  author={Wang, Furun},
  year={2003},
  school={The University of Texas at Austin},
  url = {http://hdl.handle.net/2152/28512}
}

@article{KF_ORI,
    author = {Kalman, R. E.},
    title = "{A New Approach to Linear Filtering and Prediction Problems}",
    journal = {	J. Basic Eng.},
    volume = {82},
    number = {1},
    pages = {35-45},
    year = {1960},
    month = {03},
    issn = {0021-9223},
    doi = {10.1115/1.3662552},
    url = {https://doi.org/10.1115/1.3662552}
}

@Article{COM_CALI_ZHANG,
   AUTHOR = {Zhang, Haoyue and Xu, Peng and Ye, Zongqi and Ye, Dong and Qiang, Li-E and Luo, Ziren and Qi, Keqi and Wang, Shaoxin and Cai, Zhiming and Wang, Zuolei and Lei, Jungang and Wu, Yueliang},
   TITLE = {A Systematic Approach for Inertial Sensor Calibration of Gravity Recovery Satellites and Its Application to Taiji-1 Mission},
   JOURNAL = {Remote Sensing},
   VOLUME = {15},
   YEAR = {2023},
   NUMBER = {15},
   URL = {https://doi.org/10.3390/rs15153817},
   DOI = {10.3390/rs15153817}
}

@article{COM_CALI_WEI,
  title = {Calibration of the in-orbit center-of-mass of Taiji-1},
  author = {Wei, Xiaotong and Huang, Li and Shen, Tingyang and Cai, Zhiming and He, Jibo},
  journal = {Phys. Rev. D},
  volume = {108},
  issue = {8},
  pages = {082001},
  numpages = {8},
  year = {2023},
  month = {Oct},
  publisher = {American Physical Society},
  doi = {10.1103/PhysRevD.108.082001},
  url = {https://link.aps.org/doi/10.1103/PhysRevD.108.082001}
}

@inproceedings{gath2004drag,
  title={Drag free and attitude control system design for the LISA pathfinder mission},
  author={Gath, Peter and Fichter, Walter and Kersten, Michael and Schleicher, Alexander},
  booktitle={AIAA Guidance, Navigation, and Control Conference and Exhibit},
  pages={5430},
  year={2004},
  doi={10.2514/6.2004-5430},
  url={https://doi.org/10.2514/6.2004-5430}
}

@inproceedings{gath2007drag,
  title={Drag free and attitude control system design for the LISA science mode},
  author={Gath, Peter and Schulte, Hans Reiner and Weise, Dennis and Johann, Ulrich},
  booktitle={AIAA Guidance, Navigation and Control Conference and Exhibit},
  pages={6731},
  year={2007},
  doi={10.2514/6.2007-6731},
  url={https://doi.org/10.2514/6.2007-6731}
}

@article{vidano2020lisa,
  title={The LISA DFACS: A nonlinear model for the spacecraft dynamics},
  author={Vidano, Simone and Novara, Carlo and Colangelo, Luigi and Grzymisch, Jonathan},
  journal={Aerosp. Sci. Technol.},
  volume={107},
  pages={106313},
  year={2020},
  publisher={Elsevier},
  url={https://doi.org/10.1016/j.ast.2020.106313},
  doi={10.1016/j.ast.2020.106313}
}

@article{taiji2021china,
  author={Wu, Yue-Liang and others},
  collaboration={Taiji Scientific Collaboration},
  title={China’s first step towards probing the expanding universe and the nature of gravity using a space borne gravitational wave antenna},
  journal={Commun. Phys.},
  volume={4},
  number={1},
  pages={34},
  year={2021},
  doi={10.1038/s42005-021-00529-z},
  url={https://doi.org/10.1038/s42005-021-00529-z},
  publisher={Nature Publishing Group UK London}
}

@article{LISA2024,
  title={LISA definition study report},
  author={Colpi, Monica and Danzmann, Karsten and Hewitson, Martin and Holley-Bockelmann, Kelly and Jetzer, Philippe and Nelemans, Gijs and Petiteau, Antoine and Shoemaker, David and Sopuerta, Carlos and Stebbins, Robin and others},
  journal={arXiv preprint arXiv:2402.07571},
  year={2024}
}

@article{GRS_2022,
  title={A simplified gravitational reference sensor for satellite geodesy},
  author={D{\'a}vila {\'A}lvarez, Anthony and Knudtson, Aaron and Patel, Unmil and Gleason, Joseph and Hollis, Harold and Sanjuan, Jose and Doughty, Neil and McDaniel, Glenn and Lee, Jennifer and Leitch, James and others},
  journal={Journal of Geodesy},
  volume={96},
  number={10},
  pages={70},
  year={2022},
  publisher={Springer},
  doi = {10.1007/s00190-022-01659-0}
}

@article{lisa2011lisa,
  title={LISA unveiling a hidden universe},
  author={LISA International Science Team and others},
  journal={LISA Assessment Study Report, ESA/SRE},
  volume={3},
  year={2011}
}

@phdthesis{lupi2019precise,
  title={Precise control of LISA with quantitative feedback theory},
  author={Lupi, Francesco},
  year={2019},
  school={Delft University of Technology Delft, The Netherlands}
}

@article{vidano2022lisa,
  title={The LISA DFACS: Model predictive control design for the test mass release phase},
  author={Vidano, S and Novara, C and Pagone, M and Grzymisch, J},
  journal={Acta Astronautica},
  volume={193},
  pages={731--743},
  year={2022},
  publisher={Elsevier},
  doi = {10.1016/j.actaastro.2021.12.056}
}

@phdthesis{virdis2021meteoroid,
  title={A Meteoroid Impact Recovery Control System for the LISA Gravitational Wave Observatory},
  author={Virdis, Mario},
  year={2021},
  school={Politecnico di Torino}
}

@article{ROBUST_CONTROL,
  title={Robust Control of an Input-redundant Aircraft against Atmospheric Disturbances and Actuator Faults},
  author={Trifonov, Maksim and Prochazka, Karl Frederik and Kr{\"u}ger, Saleh},
  journal={International Journal of Mechanical Engineering and Robotics Research},
  volume={8},
  number={6},
  pages={905--910},
  year={2019},
  publisher={International Journal of Mechanical Engineering and Robotics Research},
  doi = {10.18178/ijmerr.8.6.905-910}
}

@article{wang2025research,
  title={Research on Mass Center Identification for Gravitational Wave Detection Spacecraft with Guaranteed Laser Link Pointing Accuracy},
  author={Wang, Shen-Ao and Zhang, Huibo and Cai, Lin and Wang, Ziming and An, Yumin},
  journal={Remote Sensing},
  volume={17},
  number={2},
  pages={296},
  year={2025},
  publisher={MDPI},
  doi = {10.3390/rs17020296}
}

@article{OB_brzozowski2022lisa,
  title={The LISA optical bench: an overview and engineering challenges},
  author={Brzozowski, William and Robertson, David and Fitzsimons, Ewan and Ward, Henry and Keogh, Jennifer and Taylor, Alasdair and Milanova, Maria and Perreur-Lloyd, Michael and Ali, Zeshan and Earle, Andrew and others},
  journal={Space Telescopes and Instrumentation 2022: Optical, Infrared, and Millimeter Wave},
  volume={12180},
  pages={237--253},
  year={2022},
  publisher={SPIE}
}

@book{gu2005robust,
  title={Robust control design with MATLAB{\textregistered}},
  author={Gu, Da-Wei and Petkov, Petko Hristov and Konstantinov, Mihail Mihaylov},
  year={2005},
  publisher={Springer}
}

@inproceedings{akf_akhlaghi2017adaptive,
  title={Adaptive adjustment of noise covariance in Kalman filter for dynamic state estimation},
  author={Akhlaghi, Shahrokh and Zhou, Ning and Huang, Zhenyu},
  booktitle={2017 IEEE power \& energy society general meeting},
  pages={1--5},
  year={2017},
  organization={IEEE},
  doi = {10.1109/PESGM.2017.8273755}
}

@article{akf_ge2016performance,
  title={Performance analysis of the Kalman filter with mismatched noise covariances},
  author={Ge, Quanbo and Shao, Teng and Duan, Zhansheng and Wen, Chenglin},
  journal={IEEE Transactions on Automatic Control},
  volume={61},
  number={12},
  pages={4014--4019},
  year={2016},
  publisher={IEEE},
  doi = {10.1109/TAC.2016.2535158}
}

@inproceedings{akf_sage1969algorithms,
  title={Algorithms for sequential adaptive estimation of prior statistics},
  author={Sage, Andrew P and Husa, Gary W},
  booktitle={1969 IEEE Symposium on Adaptive Processes (8th) Decision and Control},
  pages={61--61},
  year={1969},
  organization={IEEE},
  doi = {10.1109/SAP.1969.269927}
}

@article{akf_ge2024adaptive,
  title={Adaptive Kalman filtering based on model parameter ratios},
  author={Ge, Quanbo and Li, Yunyu and Wang, Yuanliang and Hu, Xiaoming and Li, Hong and Sun, Changyin},
  journal={IEEE Transactions on Automatic Control},
  volume={69},
  number={9},
  pages={6230--6237},
  year={2024},
  publisher={IEEE},
  doi = {10.1109/TAC.2024.3376306}
}

@misc{hu2017taiji,
  title={The Taiji Program in Space for gravitational wave physics and the nature of gravity},
  author={Hu, Wen-Rui and Wu, Yue-Liang},
  journal={National Science Review},
  volume={4},
  number={5},
  pages={685--686},
  year={2017},
  publisher={Oxford University Press},
  doi = {10.1093/nsr/nwx116},
  url = {https://doi.org/10.1093/nsr/nwx116}
}

@article{bender2000lisa,
  title={LISA: a cornerstone mission for the observation of gravitational waves},
  author={Bender, P and others},
  journal={System and technology study report ESA-SCI},
  volume={11},
  pages={2000},
  year={2000}
}

@article{tq1_zhou2022non,
  title={Non-gravitational force measurement and correction by a precision inertial sensor of TianQin-1 satellite},
  author={Zhou, An-Nan and Cai, Lin and Xiao, Chun-Yu and Tan, Ding-Yin and Li, Hong-Yin and Bai, Yan-Zheng and Zhou, Ze-Bing and Luo, Jun},
  journal={Classical and Quantum Gravity},
  volume={39},
  number={11},
  pages={115005},
  year={2022},
  publisher={IOP Publishing},doi = {10.1088/1361-6382/ac68c9},
url = {https://doi.org/10.1088/1361-6382/ac68c9}
}

@article{stiffness_2009Current,
  title={Current LISA Spacecraft Design},
  author={ Merkowitz, S M  and  Castellucci, K E  and  Depalo, S V  and  Generie, J A  and  Maghami, P G  and  Peabody, H L },
  journal={Journal of Physics Conference Series},
  volume={154},
  number={1},
  pages={012021},
  year={2009},
  doi = {10.1088/1742-6596/154/1/012021}
}

@book{LMIboyd1994linear,
  title={Linear matrix inequalities in system and control theory},
  author={Boyd, Stephen and El Ghaoui, Laurent and Feron, Eric and Balakrishnan, Venkataramanan},
  year={1994},
  publisher={SIAM}
}

@article{DFACSbohan2024review,
  title={A review on DFACS (I): System design and dynamics modeling},
  author={Bohan, JIAO and Qifan, LIU and Zhaohui, DANG and Xiaokui, YUE and Yuanqing, XIA and Li, DUAN and Qinglei, HU and Chenglei, YUE and Pengcheng, WANG and Ming, GUO and others},
  journal={Chinese Journal of Aeronautics},
  volume={37},
  number={5},
  pages={92--119},
  year={2024},
  publisher={Elsevier},
  doi = {10.1016/j.cja.2024.01.031}
}

@ARTICLE{1972AKF_SAGEHUSA,
  author={Mehra, R.},
  journal={IEEE Transactions on Automatic Control}, 
  title={Approaches to adaptive filtering}, 
  year={1972},
  volume={17},
  number={5},
  pages={693-698},
  doi={10.1109/TAC.1972.1100100}}

@article{song2020AKF_COMATCHING,
  title={Adaptive Kalman filters for nonlinear finite element model updating},
  author={Song, Mingming and Astroza, Rodrigo and Ebrahimian, Hamed and Moaveni, Babak and Papadimitriou, Costas},
  journal={Mechanical Systems and Signal Processing},
  volume={143},
  pages={106837},
  year={2020},
  publisher={Elsevier},
  doi = {10.1016/j.ymssp.2020.106837}
}

@techreport{SRP_Georgevic1971,
  author = {Georgevic, R. M.},
  title  = {Mathematical Model of the Solar Radiation Force and Torques Acting on the Components of a Spacecraft},
  institution = {Jet Propulsion Laboratory},
  year   = {1971},
  number = {NASA TM 33-494},
  url    = {https://ntrs.nasa.gov/citations/19720004068}
}

@article{li2021orbit,
  title={Orbit determination for a space-based gravitational wave observatory},
  author={Li, Zhuo and Zheng, Jianhua},
  journal={Acta Astronautica},
  volume={185},
  pages={170--178},
  year={2021},
  publisher={Elsevier},
  doi = {10.1016/j.actaastro.2021.04.014}
}

@inproceedings{DECIGOsato2017status,
  title={The status of DECIGO},
  author={Sato, Shuichi and Kawamura, Seiji and Ando, Masaki and Nakamura, Takashi and Tsubono, Kimio and Araya, Akito and Funaki, Ikkoh and Ioka, Kunihito and Kanda, Nobuyuki and Moriwaki, Shigenori and others},
  booktitle={Journal of Physics: Conference Series},
  volume={840},
  number={1},
  pages={012010},
  year={2017},
  organization={IOP Publishing},
  doi = {10.1088/1742-6596/840/1/012010}
}

@article{TDI_tinto2021time,
  title={Time-delay interferometry},
  author={Tinto, Massimo and Dhurandhar, Sanjeev V},
  journal={Living Reviews in Relativity},
  volume={24},
  number={1},
  pages={1},
  year={2021},
  publisher={Springer}
}

@article{luo2020brief,
  title={A brief analysis to Taiji: Science and technology},
  author={Luo, Ziren and Guo, ZongKuan and Jin, Gang and Wu, Yueliang and Hu, Wenrui},
  journal={Results in Physics},
  volume={16},
  pages={102918},
  year={2020},
  publisher={Elsevier},
  doi = {10.1016/j.rinp.2019.102918}
}

@article{reigber2002champ,
  title={CHAMP mission status},
  author={Reigber, Ch and L{\"u}hr, Hermann and Schwintzer, P},
  journal={Advances in space research},
  volume={30},
  number={2},
  pages={129--134},
  year={2002},
  publisher={Elsevier},
  doi = {10.1016/S0273-1177(02)00276-4}
}

@book{GRACE_weil1997gravity,
  title={Gravity and grace},
  author={Weil, Simone},
  year={1997},
  publisher={U of Nebraska Press}
}

@article{gfo_kornfeld2019grace,
  title={GRACE-FO: the gravity recovery and climate experiment follow-on mission},
  author={Kornfeld, Richard P and Arnold, Bradford W and Gross, Michael A and Dahya, Neil T and Klipstein, William M and Gath, Peter F and Bettadpur, Srinivas},
  journal={Journal of spacecraft and rockets},
  volume={56},
  number={3},
  pages={931--951},
  year={2019},
  publisher={American Institute of Aeronautics and Astronautics},
  url = {https://arc.aiaa.org/doi/pdf/10.2514/1.A34326}
}

@article{lisataiji_wang2021alternative,
  title={Alternative lisa-taiji networks},
  author={Wang, Gang and Ni, Wei-Tou and Han, Wen-Biao and Xu, Peng and Luo, Ziren},
  journal={Physical Review D},
  volume={104},
  number={2},
  pages={024012},
  year={2021},
  publisher={APS},
  doi = {10.1103/PhysRevD.104.024012}
}

@article{liu2025design,
  title={Design of in-orbit calibration scheme for scale factor and center-of-mass deviation of inertial sensor of Taiji program for space gravitational wave detection},
  author={Liu, Chang and Wei, Xiaotong and Zhang, Haoyue and Deng, Qiong and Liang, Bo and Qiang, Li'e and Xu, Peng and Qi, Keqi and Wang, Shaoxin},
  journal={ACTA PHYSICA SINICA},
  volume={74},
  number={8},
  year={2025},
  publisher={CHINESE PHYSICAL SOC PO BOX 603, BEIJING 100080, PEOPLES R CHINA},
  doi = {10.7498/aps.74.20241376}
}

@article{LISATAIJI_ruan2020lisa,
  title={The lisa--taiji network},
  author={Ruan, Wen-Hong and Liu, Chang and Guo, Zong-Kuan and Wu, Yue-Liang and Cai, Rong-Gen},
  journal={Nature Astronomy},
  volume={4},
  number={2},
  pages={108--109},
  year={2020},
  publisher={Nature Publishing Group UK London},
  doi = {10.1038/s41550-019-1008-4}
}

@article{LISATAIJI_orlando2021measuring,
  title={Measuring parity violation in the stochastic gravitational wave background with the LISA-Taiji network},
  author={Orlando, Giorgio and Pieroni, Mauro and Ricciardone, Angelo},
  journal={Journal of Cosmology and Astroparticle Physics},
  volume={2021},
  number={03},
  pages={069},
  year={2021},
  publisher={IOP Publishing},
  doi = {10.1088/1475-7516/2021/03/069}
}

@article{lisataiji_cornish_shuman2022massive,
  title={Massive black hole binaries and where to find them with dual detector networks},
  author={Shuman, Kevin J and Cornish, Neil J},
  journal={Physical Review D},
  volume={105},
  number={6},
  pages={064055},
  year={2022},
  publisher={APS},
  doi = {10.1103/PhysRevD.105.064055}
}

@article{DECIGOkawamura2011japanese,
  title={The Japanese space gravitational wave antenna: DECIGO},
  author={Kawamura, Seiji and Ando, Masaki and Seto, Naoki and Sato, Shuichi and Nakamura, Takashi and Tsubono, Kimio and Kanda, Nobuyuki and Tanaka, Takahiro and Yokoyama, Jun'ichi and Funaki, Ikkoh and others},
  journal={Classical and Quantum Gravity},
  volume={28},
  number={9},
  pages={094011},
  year={2011},
  publisher={IOP Publishing},
  doi = {10.1088/0264-9381/28/9/094011}
}

@article{DECIGOando2010decigo,
  title={DECIGO and DECIGO pathfinder},
  author={Ando, Masaki and Kawamura, Seiji and Seto, Naoki and Sato, Shuichi and Nakamura, Takashi and Tsubono, Kimio and Takashima, Takeshi and Funaki, Ikkoh and Numata, Kenji and Kanda, Nobuyuki and others},
  journal={Classical and Quantum Gravity},
  volume={27},
  number={8},
  pages={084010},
  year={2010},
  publisher={IOP Publishing},
  doi = {10.1088/0264-9381/27/8/084010}
}

@article{TQ_liu2020science,
  title={Science with the TianQin observatory: Preliminary results on stellar-mass binary black holes},
  author={Liu, Shuai and Hu, Yi-Ming and Zhang, Jian-dong and Mei, Jianwei},
  journal={Physical Review D},
  volume={101},
  number={10},
  pages={103027},
  year={2020},
  publisher={APS},
  doi = {10.1103/PhysRevD.101.103027}
}

@article{TQ_mei2021tianqin,
  title={The TianQin project: Current progress on science and technology},
  author={Mei, Jianwei and Bai, Yan-Zheng and Bao, Jiahui and Barausse, Enrico and Cai, Lin and Canuto, Enrico and Cao, Bin and Chen, Wei-Ming and Chen, Yu and Ding, Yan-Wei and others},
  journal={Progress of Theoretical and Experimental Physics},
  volume={2021},
  number={5},
  pages={05A107},
  year={2021},
  publisher={Oxford University Press},
  doi = {10.1093/ptep/ptaa165},
  url = {https://doi.org/10.1093/ptep/ptaa165}
}

@article{ni2002astrod,
  title={ASTROD--an overview},
  author={Ni, Wei-Tou},
  journal={International Journal of Modern Physics D},
  volume={11},
  number={07},
  pages={947--962},
  year={2002},
  publisher={World Scientific},
  doi = {10.1142/S0218271802002499},
  url = {https://www.worldscientific.com/doi/abs/10.1142/S0218271802002499}
}

@article{ni2008astrod,
  title={ASTROD and ASTROD I—overview and progress},
  author={Ni, Wei-Tou},
  journal={International Journal of Modern Physics D},
  volume={17},
  number={07},
  pages={921--940},
  year={2008},
  publisher={World Scientific},
  doi = {10.1142/S0218271808012619}
}

@article{caron1997virgo,
  title={The virgo interferometer},
  author={Caron, Bernard and Dominjon, A and Drezen, C and Flaminio, R and Grave, X and Marion, F and Massonnet, L and Mehmel, C and Morand, R and Mours, B and others},
  journal={Classical and Quantum Gravity},
  volume={14},
  number={6},
  pages={1461},
  year={1997},
  publisher={IOP Publishing}
}

@article{kagra_sekiguchi2012current,
  title={Current status of numerical-relativity simulations in Kyoto},
  author={Sekiguchi, Yuichiro and Kiuchi, Kenta and Kyutoku, Koutarou and Shibata, Masaru},
  journal={Progress of Theoretical and Experimental Physics},
  volume={2012},
  number={1},
  pages={01A304},
  year={2012},
  publisher={Oxford University Press}
}

@article{lisa_virgo_kagra_abbott2020prospects,
  title={Prospects for observing and localizing gravitational-wave transients with Advanced LIGO, Advanced Virgo and KAGRA},
  author={Abbott, Benjamin P and Abbott, R and Abbott, TD and Abraham, Sheelu and Acernese, Fausto and Ackley, K and Adams, C and Adya, VB and Affeldt, C and Agathos, M and others},
  journal={Living reviews in relativity},
  volume={23},
  number={1},
  pages={3},
  year={2020},
  publisher={Springer},
  doi = {10.1007/s41114-020-00026-9}
}

@article{xu2024finite,
  title={Finite frequency domain {$H_{\infty}$} hybrid control design of drag-free spacecraft with model-based generalized extended state observer},
  author={Xu, Qianjiao and Cui, Bing and Wang, Pengcheng and Xia, Yuanqing and Zhang, Yonghe},
  journal={Control Engineering Practice},
  volume={153},
  pages={106096},
  year={2024},
  publisher={Elsevier},
  doi = {10.1016/j.conengprac.2024.106096}
}

@article{yidi2023robust,
  title={Robust controller design for drag-free satellites with two test masses},
  author={Yidi, FAN and Pengcheng, WANG and Wei, LU and Ke, AN and Yonghe, ZHANG},
  journal={Journal of Deep Space Exploration},
  volume={10},
  number={3},
  pages={310--321},
  year={2023},
  publisher={Beijing Institute of Technology Press}
}

@article{diaz2013design,
  title={Design of the magnetic diagnostics unit onboard LISA Pathfinder},
  author={Diaz-Aguil{\'o}, Marc and Mateos, Ignacio and Ramos-Castro, Juan and Lobo, Alberto and Garc{\'\i}a-Berro, Enrique},
  journal={Aerospace science and technology},
  volume={26},
  number={1},
  pages={53--59},
  year={2013},
  publisher={Elsevier}
}

\end{document}